\documentclass[letterpaper,11pt]{JHEP3}
\usepackage{epsfig, array, cite, latexsym, amsmath, oldgerm}
\usepackage{verbatim}  
\usepackage[all]{xy}
\preprint{SLAC-PUB-13386}
\newcommand\eqn[1]{\label{eq:#1}} 
\newcommand\Eq[1]{eq.~\eqref{eq:#1}} 
\newcommand\half{{\textstyle{\frac{1}{2}}}} 
\newcommand\fourth{{\textstyle{\frac{1}{4}}}} 
\newcommand\fourthi{{\textstyle{\frac{i}{4}}}}

\newcommand\bfz{\mathbf{Z}}

\newcommand{\beq}{\begin{eqnarray}}
\newcommand{\eeq}{\end{eqnarray}}

\newcommand{\CD}{{\cal D}}
  
\newcommand{\CF}{{\cal F}}

\newcommand{\CN}{{\cal N}}

\newcommand{\CQ}{{\cal Q}}

\newcommand{\CL}{{\cal L}}
\newcommand{\bfn}{{\bf n}}

\newcommand{\bfe}{{\bf  e}}
\newcommand\bfmu{\boldsymbol{\mu}}
\newcommand\bfZ{\mathbf{Z}}

\newcommand{\R}{\mathbb{R}}
\DeclareMathOperator{\Tr}{Tr}

\newcommand{\sla}[1]%
        {\kern .25em\raise.18ex\hbox{$/$}\kern-.55em #1}

\def\Dslash{{\rlap{\raise 1pt \hbox{$\>/$}}D}}
\newcommand{\mybar}[1]%
        {\kern 0.6pt\overline{\kern -0.6pt#1\kern -0.6pt}\kern 0.6pt}
\newcommand{\dig}{\kern-1.5pt \raisebox{.9ex}{$\cdot$}  \kern1.5pt
  \raisebox{0ex}{${\mathbf\cdot}$}\kern1.5pt \raisebox{-.9ex}{$\cdot$}} 
\newcommand{\digb}{\kern-1.5pt \raisebox{.75ex}{$\cdot$}  \kern1.5pt
  \raisebox{0ex}{${\mathbf\cdot}$}\kern1.5pt \raisebox{-.75ex}{$\cdot$}} 
\newcommand{\digc}{\kern-1.5pt \raisebox{1.05ex}{$\cdot$}  \kern1.5pt
  \raisebox{0ex}{${\mathbf\cdot}$}\kern1.5pt \raisebox{-1.05ex}{$\cdot$}} 
\newcommand\fverb{\setbox\pippobox=\hbox\bgroup\verb}
\newcommand\fverbdo{\egroup\medskip\noindent%
                        \fbox{\unhbox\pippobox}\ }
\newcommand\fverbit{\egroup\item[\fbox{\unhbox\pippobox}]}
\newbox\pippobox

\title{Deformed matrix models,  supersymmetric lattice twists and 
$\CN=\fourth$ supersymmetry
}

\author
    {
    {
      \def\href#1#2{#2}	
    Mithat \"Unsal$^1$\footnote{\email{unsal@slac.stanford.edu}}
    \\${}^1$ SLAC and Physics Department, Stanford University, Stanford, CA 94305
       }
     }%


\abstract{
A  manifestly supersymmetric nonperturbative matrix regularization for  a twisted version of 
 $\CN=(8,8) $ theory on a curved background (a two-sphere) is constructed.  
 Both continuum and the matrix regularization respect 
 four exact  scalar supersymmetries under a twisted version of the 
 supersymmetry algebra.  
 We then discuss   a succinct $\CQ=1$ deformed matrix model regularization of $ \CN=4$  SYM 
 in $d=4$, which is equivalent to a non-commutative   $A_4^*$ orbifold lattice formulation.  
  Motivated by recent progress in supersymmetric  lattices,   we also propose a    $ \CN=\fourth$   supersymmetry preserving deformation of 
    $ \CN=4$  SYM theory on $\R^4$.
    In   this class of   $ \CN=\fourth$   theories,  
  both the regularized   and continuum theory respect the same set of (scalar) supersymmetry. 
   By using the equivalence of the deformed matrix models with the lattice formulations, we give a very simple physical argument on why the exact lattice supersymmetry must be a subset of scalar subalgebra. This argument  disagrees with  the   recent claims  of the  link approach, for which    we  give a new  interpretation. 
}

\begin{document}

\section{Introduction} 
\label{sec:intro}
This paper has three goals: One is to construct  a manifestly supersymmetric matrix  (non-lattice) regularization  for certain twisted supersymmetric gauge theories formulated on curved backgrounds, such as $S^2$ or $S^2 \times \mathbb R$.  The other purpose is to discuss the global supersymmetry in the context of twisted supersymmetry,  deformed (supersymmetric) matrix models,  supersymmetric lattices, and supersymmetry in curved spaces.  We hope to 
provide a   sharp meaning to the notion of exact lattice supersymmetry  by doing this. 
Our last goal is to introduce a simpler  deformed matrix model 
regularization for $\CN=4$ supersymmetric Yang-Mills (SYM)  theory  
in $d=4$ dimensions  and discuss  its relation to the supersymmetric 
lattice regularization. 

It is well-known that  global  scalar supersymmetry may be carried  to curved spaces if a 
twisted version of the supersymmetry algebra is used \cite{Witten:1988ze}.   On a flat space, 
twisting is a procedure which embeds a new  
Lorentz  group into the product of the usual Lorentz and a global symmetry group. Usually, this 
is done in such a way that  some of the spinors of the Lorentz symmetry turns into 
 spin-0  scalars under the new twisted Lorentz group, see for example  
 \cite{Kato:2005fj}.   
 The twisted theories can be carried into curved backgrounds while preserving 
 the  (nilpotent) scalar supersymmetry generators, or the scalar subalgebra. 
 A subclass of twisted theories which admits scalar supercharge 
may also be defined on lattices without upsetting the scalar subalgebra. 
  (Not all twisted theories with a nilpotent supercharge admit a  lattice regularization, see the discussion in \S\ref{sec:SL})  
 We refer to this subclass as {\bf supersymmetric lattice twists} or {\bf SL-twists} for short. 
The  existence of a  nilpotent scalar supersymmetry  $Q^2 =0$  is  sufficient to formulate 
a  topologically twisted version of supersymmetric gauge theories on curved  spaces. 
The same criterion, however,  is necessary but not sufficient to construct a physical 
 (non-topological)  supersymmetric theory  on a  lattice. 

Motivated by these general  observations,  we first construct a deformed supersymmetric matrix model  regularization for a twisted theory on curved background,  a  two-sphere $S^2$.  The remarkable 
property of this construction is that both the regularized theory and continuum theory respect
the same set of scalar supersymmetries. 
Our target theory is  a twisted version (which we refer as A-twist) of  $\CN= (8,8)$  SYM  theory with gauge group $U(k)$  residing on a  two-sphere, $S^2$.  Both the deformed matrix model and the A-twist  has  $\CQ=4$  scalar  supersymmetries 
and these are the exact supersymmetries of the target theory on $S^2$, with no enhancement of supersymmetry in the continuum limit.

 Next, we study a $\beta$ (flux) deformation of the Type IIB matrix model.\footnote{This model is essentially the Leigh-Strassler deformation of 
$\CN=4$  SYM theory in $d=4$ reduced to a matrix model  \cite{Leigh:1995ep}.  The $\beta$-deformed model, without any orbifold projections, serves as a non-perturbative regulator for the 
target $\CN=(8,8)$ theory. The model 
 was studied 
in \cite{Nishimura:2003tf,Unsal:2004cf}, however, the unnecessity of  orbifolding and the emergence of the base space from the zero-action configurations was  recognized 
 later \cite{Unsal:2005us}.}  
The continuum limit of this model is $\CN= (8,8)$ SYM theory on flat torus, $T^2$.  The regularized matrix model 
has $\CQ=4$ scalar supersymmetries, which are the scalar set of supersymmetries of a B-twist 
of   $\CN= (8,8)$ target theory.  In the continuum limit, the supersymmetry enhances to the full 
sixteen supersymmetries.
A more interesting case is a  certain two-flux deformation  of  the matrix model. 
This deformation  preserves only $\CQ=1$ out of   $\CQ=16$ supersymmetries, and generates the   $\CN=4$ SYM on $T^4$ in its classical continuum limit.   We will benefit 
 from the relation of the deformed matrix models and supersymmetric lattices 
  in the discussion of the exact global supersymmetries of various lattice formulations.  
 
In recent years, there has been significant progress on the non-perturbative lattice construction 
for the supersymmetric gauge theories.  Various approaches are used to construct supersymmetric lattices with exact supersymmetry at finite lattice spacing.  \footnote{Alternatives formulations in which supersymmetry only  emerges in the continuum 
are studied 
in, for example, \cite{Suzuki:2005dx, Elliott:2008jp, lee-2007-76}. Also see the interpretation in 
\S\ref{link}.} One such approach is the orbifold constructions  which preserve a nilpotent subset of  supersymmetries \cite{Kaplan:2002wv,  Cohen:2003xe, Cohen:2003qw}.  Also see \cite{
 Endres:2006ic, Onogi:2005cz, Ohta:2006qz,Giedt:2003xr,Giedt:2006pd,Unsal:2005yh} for related work.
Catterall   \cite{Catterall:2003wd,Catterall:2004np,Catterall:2005fd,Catterall:2006jw } and Sugino \cite{Sugino:2003yb,Sugino:2004qd,Sugino:2004uv,Sugino:2006uf,Sugino:2008yp}, starting with a  ``topologically" twisted form of the target theories, successfully preserved a scalar (nilpotent) subset of supersymmetries on the lattice.   The relation between these three formulations  was not clear at first. 

\begin{figure}
\begin{center}
\includegraphics[angle=-90,width=4 in]{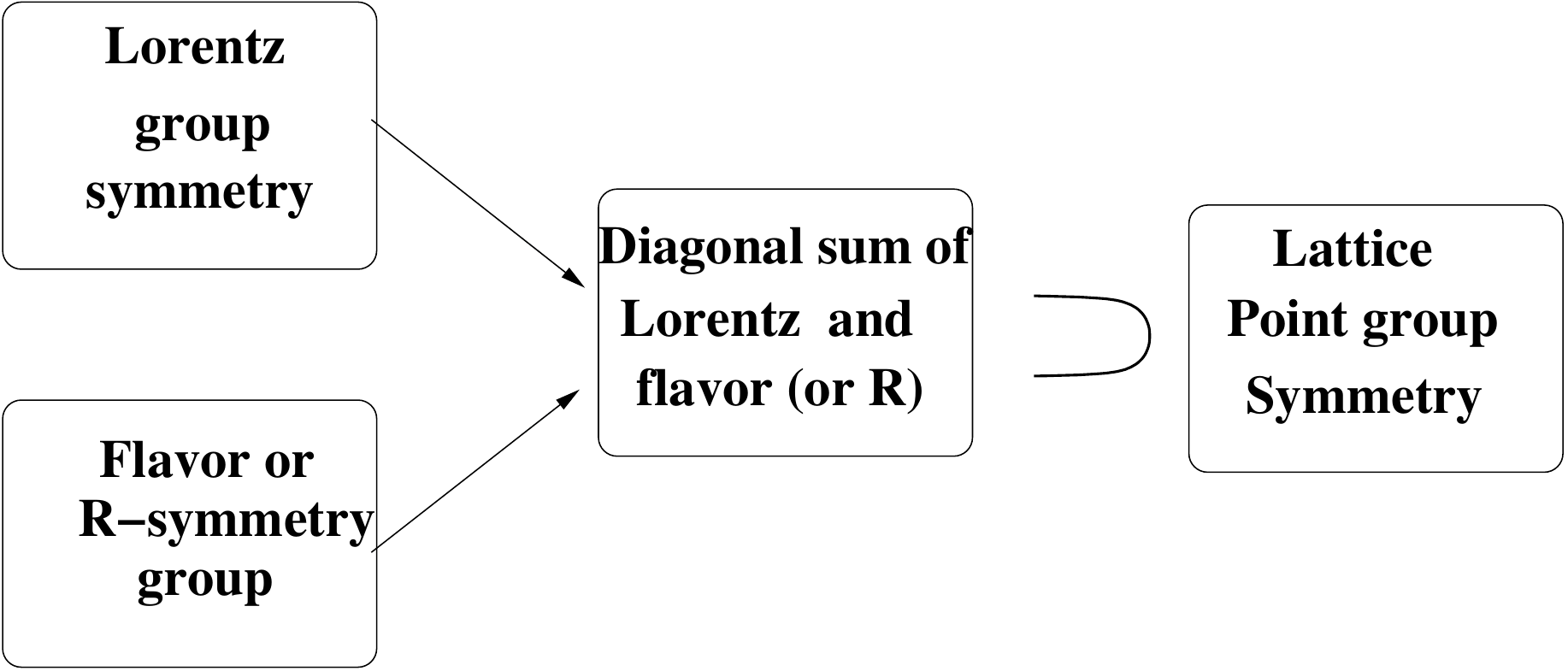}
\caption
    {\small
 Twisting is one of the the  main ideas in the formulation of  {\bf i)} supersymmetric lattices for physical 
 SYM theories,   {\bf ii)}  the staggered fermion formulation of lattice QCD,  and  {\bf iii)} in the formulation of   topological versions of SYM theories  on curved spaces.
 [In literature,  twisting is sometimes referred as ``topological twisting" due to its applications to topological theories. This  is a  misnomer.]
    In  neither of the first  two cases (which is the main interest of this paper),  
 the lattice point group symmetry may be considered as a sub-group of Lorentz symmetry. 
 It is in fact a subgroup of the diagonal sum of the Lorentz and $G_{\rm global}$, referred as twisted Lorentz  group $SO(4)'$. 
 $G_{\rm global}$ is an ${\cal R}$-symmetry in the case of supersymmetric theories and is a 
 flavor symmetry in the case of QCD.  In  both lattice formulations,  
 the fermions are in single-valued, integer spin representation of $SO(4)'$ and its discrete subgroups. Of course, by undoing the twist, we recover usual fermions   with double-valued spinor representation under $SO(4)$. The figure is adapted from Ref.\cite{Kaplan:2007zz}.
 }
\label{godd}
\end{center}
\end{figure}

Motivated by the unconventional aspects of  supersymmetric orbifold lattices, such as scalars of the target theories residing on the links (rather than on sites) and fermions filling single-valued integer spin representations
(rather than being  double-valued spinors), Ref.\cite{Unsal:2006qp} showed that   all the supersymmetric orbifold lattices do indeed produce a twisted version of the supersymmetric 
gauge theories in their  continuum.  The main point of Ref.\cite{Unsal:2006qp} is depicted in 
Fig.~\ref{godd}.
This observation, merged the  ``topological" approach  and  supersymmetric orbifold lattices at the conceptual level. Soon after,  Catterall \cite{Catterall:2007kn} showed that, the use of the 
correct  twist  together with the geometrical discretization
rules produce the supersymmetric lattice actions for the orbifold lattices. 
 More recent important work by Takimi \cite{Takimi:2007nn}, and Damgaard et.al. \cite{Damgaard:2007xi,Damgaard:2007eh,Damgaard:2008pa} 
demonstrated  the equivalence of these lattice formulations even at finite lattice spacing. 
Ref.\cite{Damgaard:2007be}  also provided a full classification of the supersymmetric lattices 
that can be obtained by orbifolding and argued for uniqueness in certain cases. 
  
There is one other approach to lattice supersymmetry which aims to preserve { \it all } the supersymmetries on the lattice, not only the nil-potent scalar supercharges. This approach is also motivated by the twisted form of the supersymmetry algebra and Dirac-K\"ahler structure of the fermions. \footnote{An earlier proposal of using   Dirac-K\"ahler fermions to supersymmetric lattices appeared in 
\cite{Elitzur:1982vh}.}
It is  referred as   {\it link approach} in D'Adda et.al.
\cite{D'Adda:2004jb,D'Adda:2004ia, D'Adda:2005zk,D'Adda:2007ax}. 
The  claim of preserving the whole set of supersymmetries is  debated in 
Ref.\cite{Bruckmann:2006ub,Bruckmann:2006kb}, with a negative conclusion. 
On the other hand, 
the lattice structures on the  link approach can   be obtained by orbifold projections 
 and in fact, these two approaches are also equivalent as shown in  \cite{Damgaard:2007eh}.
 However, some lattices of link approach constructions, which are claimed to possess all the supersymmetries, can be obtained by orbifold projections which preserve either no supersymmetry, or just few scalar supercharges according to the criteria of 
  \cite{Kaplan:2002wv, Kaplan:2005ta}.   In our opinion,   Ref. \cite{Damgaard:2007eh}
 answered this question satisfactorily by showing that whatever supersymmetry remains intact under the orbifold projection is indeed the exact supersymmetry on the lattice. However, 
 \cite{Arianos:2007nv, Arianos:2008ai}  asserted the consistency of the link approach despite 
 Ref.\cite{Bruckmann:2006ub,Bruckmann:2006kb,Damgaard:2007eh}.
 Here, we will  give an independent and  simpler argument 
 which shows that  the amount of global supersymmetries  preserved in lattice regularization of the extended supersymmetric gauge theories 
 is always  the subset of {\it scalar supersymmetries}, and not the whole set of supersymmetry.

  The merit of our argument is in its conceptual simplicity. We will benefit from the (supersymmetric) deformed matrix models. Recall that  in $d=4$ dimensions, the   $\beta$ 
 \cite{Leigh:1995ep}   or a mass deformation  of superpotential reduce the $\CN=4$  SYM down to $\CN=1$.  There is no ambiguity in the amount of global supersymmetry here, 
  because  the other twelve supersymmetries are explicitly spoiled by the deformation.   We will consider similar $\CQ=1$ \cite{Unsal:2005us} and $\CQ=4$ supersymmetry preserving deformation of  $\CQ=16$   Type IIB model. The equivalence of the deformed supersymmetric matrix model to 
 supersymmetric lattices is explicitly given in \cite{Unsal:2005us} and is summarized 
 in \S\ref{matrixm}. 
   The reason that I choose to  go through this {\it detour} is to avoid the  technical 
   discussion of  the 
  modified Leibniz rule or its spin-offs such as  ``link supercharge"  altogether. 
    Our argument is very simple.  Since the  $\beta$-deformed   matrix model is equivalent to the lattice regularization, 
the existence of the   full set of supersymmetry in the lattice formulation would have 
implied that the $\beta$-deformation does not reduce the amount of  supersymmetry, which is a contradiction. This   argument also  leads us to the conclusion that exact lattice supersymmetry must be a subset of the scalar supersymmetry subalgebra.

The question of whether we can preserve the whole set of supersymmetries of a continuum gauge theory in its lattice (or matrix) regularization leads us to a surprising 
reverse-engineering.  There exist deformations of  $\CN=4$ SYM theory on $\R^4$ or $T^4$ 
which preserve  only  $\CN=\fourth$ (or $\CQ=1$) supersymmetry. The idea is to deform the twisted action such that 
only scalar super-charge remains as a global supersymmetry.  Although these theories 
looks like a BRST-like gauge fixing of an underlying gauge invariant theory, 
I was unable to construct  them  this way.  A  physical  interpretation for the  $\CN=\fourth$  theories on flat space  is  currently lacking, although their non-perturbative 
lattice and matrix regularizations    exist, and are given here.  The generalization of these 
twisted-deformed actions  to other dimensions is obvious. 

Finally, I discuss twisting in a more general context (outside supersymmetry or topological twisting)  in \S\ref{sec:twisting}  where  I rephrase 
the  staggered  and  reduced staggered fermions of lattice QCD as  particularly elegant  
applications  of the twisting idea, see Fig.\ref{godd}.   
 In these cases, the ${\cal R}$-symmetry is replaced by the flavor symmetry  of QCD.  

\section{Target theories: Twisted  $\CN=(8,8) $ SYM theories on $S^2$ and $T^2$}
\label{sec:twistedE}
Our two dimensional target theories  are  the  twisted versions of  the $\CN=(8,8)$ (or $\CQ=16$)  supersymmetric Yang-Mills theory formulated on a two-sphere, $S^2$ and on a two-torus $T^2$. 
The $\CN = (8,8)  $ theory on $\R^2$ can be obtained as the dimensional reduction of the 
$\CN=1 $ gauge theory on $\R^{10}$ down to $\R^2$. The ten dimensional theory possess an 
$SO(10)$ Euclidean Lorentz rotation group. Upon reduction,  $SO(10)$ group decompose into 
 \beq
SO(10) \longrightarrow 
\left( \begin{array}{c|c}
     SO(2) & \\
\hline     
 & SO(8)
       \end{array} \right)
\eeq
where $SO(2)$  is the two  dimensional Lorentz symmetry acting  
on $\R^2$ and $SO(8)_{\cal R} $ is the internal ${\cal R}$-symmetry group. 

We will consider  a compactification of the  $\CQ=16$ SYM theory on two-sphere, 
$S^2$. A straight forward compactification of a supersymmetric theory on a 
curved manifold breaks all the supersymmetries, since there are no
covariantly constant spinors  on curved spaces. This  can be evaded by using
a twisting procedure which turns some of the (spinor) supersymmetries  into scalars under a new Lorentz group.   The scalar supersymmetries are globally 
defined even when the underlying manifold is curved. The other supersymmetries of the flat-space  theory are no longer symmetries in the global sense on $S^2$.

The fermions and supercharges transform under the spin group of $SO(2)\times SO(8)$ as 
$( \half,
{\bf 8}) \oplus ( - \half, {\bf 8'})$ and the scalars and gauge bosons 
 transform as 
$(1, {\bf 8_v})$ and $(\pm 1, {\bf 1})$. \footnote{Under the $SO(2)$,
  the gauge boson is in two dimensional vector representation. 
Under $Spin(2)_E=U(1)_E$, it splits into two one dimensional representations.}
Below, we wish to examine two twists  of the theory that will accommodate four scalar supersymmetries. 


\subsection{A-twist}
\label{sec:twist1}
The main idea of twisting is to embed a new rotation group into the product of 
Lorentz and ${\cal R}$-symmetry groups in such a way that a subset of supercharges transform as scalars.  
 In our  particular case, we find an embedding 
of  $SO(2)'_E$ into     $SO(2)_E\times SO(8)_{\cal R}$.   Let us 
decompose $SO(8)_{\cal R} \rightarrow SO(4) \times SO(4) \sim SU(2)_A \times 
SU(2)_B \times SU(2)_L \times SU(2)_R$. Therefore, under 
\begin{equation}
SO(2)_E \times [SU(2)]^4  
\end{equation}   
the fermions and bosons branch   as 
\beq 
&& {\rm scalars }:\;\; (0, {\bf 2,2, 1,1}) \oplus   (0, {\bf 1,1,2,2}) \qquad 
{\rm gauge \; bosons}:\;\;  (\pm 1, {\bf 1,1,1,1}) \cr \cr
 &&  {\rm fermions: } \;\; ( \half,  {\bf 2,1, 2,1})  \oplus 
(-\half,  {\bf 2,1, 1,2})   
 \oplus  ( \half,  {\bf 1,2, 2,1})  \oplus  
(- \half,  {\bf 1,2, 1,2})
\eeq
One can use,  the $U(1)_A$  subgroup of the  $SU(2)_A$  to construct a twisted rotation  
group 
\beq
U(1)'_E= {\rm Diag}( U(1)_E \times U(1)_A) \qquad .
\eeq   
and declare $U(1)'_E \sim SO(2)'_E$ as the new  Lorentz  group.  This is the procedure of twisting. 
Since the $[SU(2)]^3$ subgroup 
of the full  ${\cal R}$-symmetry remains intact, it will be  an   ${\cal R}$-symmetry group of the twisted version. 
Under  $U(1)'_E \times [SU(2)]^3 $, the transformation properties of fields are 
\beq 
&&{  \rm fermions} \longrightarrow   ( 1,  {\bf 1, 2,1})   \oplus   ( 0,  {\bf 1, 2,1}) 
\oplus   ( -1,  {\bf 1, 1,2}) 
  \oplus  ( 0,  {\bf 1, 1,2})  
\oplus    ( \half,  {\bf 2, 2, 1})  \oplus  
  (- \half,  {\bf 2, 1,2}) \qquad \cr\cr
 && {\rm scalars} \longrightarrow  (\pm \half, {\bf 2, 1,1})  \oplus  
 (0, {\bf 1,2,2}) 
\qquad 
{\rm gauge \; bosons}  \longrightarrow (\pm 1, {\bf 1,1,1}) 
\eqn{twist-1}
\eeq 
Four of the fermionic degrees  of freedom  are neutral under 
the twisted
Lorentz group   $U(1)'_E$, and therefore they transform as scalars. The
same argument is also true for the supercharges, and consequently 
the sixteen supercharges of the original theory decompose precisely as fermions in 
\Eq{twist-1}.
Four of them are scalars 
\beq 
{\rm Scalar \; \;  supercharges :}
        ( 0,  {\bf 1, 2,1})  
 \oplus  ( 0,  {\bf 1, 1,2})  
\eqn{scalar-1}
\eeq 
and they can be defined    globally on curved  two-manifolds. 

There is a natural supermultiplet structure that can be read-off from the transformation  properties of the fields and supercharges:
\beq
\begin{array}{|c|c|c|}
\hline 
& {\rm Bosons} & {\rm Fermions} \cr
\hline
U(1)'_E\;\;\;  {\rm scalars}:\;\;&  (0, {\bf 1,2,2}) \;\;  & ( 0,  {\bf 1, 2,1}) \oplus 
\;\; ( 0,  {\bf 1, 1,2}) 
\qquad  \cr 
& &\cr
\hline
U(1)'_E\;\;\; {\rm spinors}:\;\; & (\pm \half, {\bf 2, 1,1}),  &\;\; ( \half,  {\bf 2, 2, 1}), \;\;
(- \half,  {\bf 2, 1,2}) \cr 
& & \cr 
\hline
 U(1)'_E \; \;\;{\rm vectors }:\;\; &  (\pm 1, {\bf 1,1,1}), & \;\;( 1,  {\bf 1, 2,1}), \;\; 
( -1,  {\bf 1, 1,2}) \;\;\;\;\;\; \cr 
& &  \cr
\hline
\end{array}
\eqn{twistf}
\eeq
The supermultiplets transform as their lowest components  and are 
respectively  scalars, spinors and vectors under   $U(1)'_E$: Unlike the supersymmetric lattice twists which associate  all the degrees of freedom with integer valued representations of the 
twisted rotation group, this twist has double-valued spinor representations as well. 

The A-twist  of the gauge theory on $S^2$ arises naturally out of a    $\CQ=4$ supersymmetry preserving mass  deformation of  $\CQ=16$ type IIB matrix model, as it will be discussed in \S\ref{sec:spectrum}.

\subsection{B-twist } 
\label{sec:twist2}
The  supersymmetric  lattice twists  have   no half-integer representation under the twisted group.  We can build   such a twist 
starting with  \Eq{twist-1} and by taking the diagonal sum of the  
  $U(1)'_E$ with the $U(1)_B$ subgroup of $SU(2)_B$: 
\beq
U(1)''_E= {\rm Diag}( U(1)'_E \times U(1)_B)=  {\rm Diag}( U(1)_E\times 
U(1)_A \times U(1)_B) 
\eeq
  This amounts to defining a charge $Q_E''$ as 
\beq 
Q''_E= Q_E + Q_A +  Q_B
\eeq
under $SO(2)_E'' \sim U(1)''_E$. 
Under $U(1)''_E \times SU(2)_L \times SU(2)_R$, we have  
\beq 
&& {\rm scalars} \longrightarrow (\pm 1, {\bf 1,1}) \oplus   2 (0, {\bf 1,1}) \oplus   
(0, {\bf 2,2}) 
\qquad 
{\rm gauge \; bosons}  \longrightarrow  (\pm 1, {\bf 1,1}) \cr \cr
&&  \rm fermions \longrightarrow 2 \times \left[( 1,  {\bf  2,1})  \oplus   ( 0,  {\bf  2,1})  
\oplus
( -1,  {\bf  1,2})  \oplus  ( 0,  {\bf  1,2})  \right]
\eqn{twist-2}
\eeq  
This twist is the one which emerges naturally from  the $A_2^*$ hexagonal lattice construction  \cite{Unsal:2006qp}.  In \S\ref{Sec:B} we will see that the B-twist also appears in a   $\CQ=4$ supersymmetry preserving $\beta$-flux deformation of  type IIB matrix model. 

\section{Matrix regularization of target theories on $S^2$ and $T^2$}
\label{SusySphere}
In this section, we 
present a manifestly supersymmetric  deformed matrix model regularization 
for the twisted version 
of supersymmetric gauge theory on $S^2$ and $T^2$.   The two types of matrix regularizations have manifest $\CQ=4$ supersymmetries, and in their continuum, correspond to  A-twist  and B-twist, respectively. 
On $S^2$, there is no enhancement of the supersymmetry in the continuum.  For $T^2$,  as in the supersymmetric  lattices, generically a nilpotent scalar subset of supersymmetry is  preserved exactly on the regularized theory,  and the others emerge accidentally in the continuum.

\subsection{The  Type IIB matrix model in $\CQ=4$ multiplets}
\label{sec:action}
To describe our regularization scheme, it is convenient to express the 
 $\CQ=16$ Type IIB matrix  theory  in 
a manifestly 
$\CQ=4$ supersymmetric  formalism.  This is most easily done by writing the 
 $\CN=4$ SYM in $d=4$ 
dimensions using  $\CN=1$ superfields followed by  dimensional reduction 
 down to $d=0$ dimension. 
The matrix  model  action is 
\beq
\label{eq:action}
S &=& \frac{1}{g^2} 
\Tr \biggl[ 
\int d^2\theta \; d^2 \mybar \theta \;\; \mybar {\bfZ}_m 
e^{2{\bf V}} {\bfZ}^m  e^{-2 {\bf V}}+ \; 
\frac{1}{4}\int d^2  \theta \;{\bf W}^{\alpha}{\bf W}_{\alpha} +  {\rm a.h.} 
 \\ \nonumber 
&+& \frac{\sqrt 2 }{3!} \epsilon_{mnp}  \int d^2 \theta  \, {\bfZ}^m 
[\bfZ^n, \bfZ^p]   +  {\rm a.h.} 
\biggr]
\eeq
Here, $\bfZ^m$, $\mybar \bfZ_m$, ${\bf V}$ and  ${\bf W}_{\alpha} $ are the dimensional 
reduction of familiar 
 $\CN=1$ chiral, anti-chiral vector and field-strength supermultiplets on $\R^4$
down to $d=0$ dimension. 

The type IIB matrix model possesses  a global $SO(10)$ symmetry and  sixteen
supersymmetries.  The $\CQ=4$ superfield language only makes the 
\beq 
SU(3)\times SU(2)\times SU(2)\times U(1)  
\eqn{decom}
\eeq
subgroup  manifest.
 The ${\bf 10}$   bosons and 
${\bf 16}$ 
fermions of $SO(10)$ decompose under  \Eq{decom} as 
\beq
{\bf 10} \ \longrightarrow\  z^m \oplus \mybar z_m \oplus  
\tilde v
\sim ({\bf 3,1,1})_{\frac { 2}{ 3} } \oplus  
({\bf \mybar 3,1,1})_{- \frac{ 2} { 3}  } 
\oplus  ({\bf 1,2,2})_{0} \ .
\eeq
\beq
{\bf 16} \ 
\longrightarrow\   \psi^m \oplus \mybar \psi_m  \oplus \lambda   \oplus  
\mybar  \lambda \sim 
({\bf 3,2,1})_{-\frac { 1}{ {3}}} \oplus 
({\bf \mybar 3,1,2})_{\frac { 1}{ { 3}}}
\oplus  ({\bf 1,2,1})_{ { 1} }
\oplus  ({\bf 1,1,2})_{ -{ 1}}\ \ 
\eeq

Next,   we construct a  $\CQ=4$ supersymmetry preserving 
mass deformation of  the matrix model
\Eq{action}.  The deformed matrix model, around a particular background solution,  
produces a higher dimensional twisted supersymmetric gauge field theory. 


\subsection{Mass deformed matrix model and symmetries}
The $ \CQ=4 $ supersymmetry preserving (equal) mass deformation of the $\CQ=16$ supercharge theory  is the dimensional reduction of the  ${\cal N}=1^* $ SYM from $d=4$  down to  $d=0$ dimension.  The ${\cal N}=1^* $ deformation of the ${\cal N}=4 $ was  studied in 
Ref.~\cite{Vafa:1994tf}.  The  deformed ``superpotential" is 
\beq
\eqn{deformation}
W( {\bfZ}^m) = \frac{\sqrt 2}{3!} \epsilon_{mnp}  \, {\bfZ}^m  [\;  \bfZ^n, 
\bfZ^p\;]   +   \frac{ 1}{\sqrt 2} \;  m \sum_{p=1}^{3}   \; \; (\bfZ_p)^2 
\eeq 
or equivalently, the deformed action is 
\beq
\eqn{deformation2}
S^{\rm deformed} = S+  \frac{ 1}{\sqrt 2}   m  \int d^2 \theta \;   \sum_{p=1}^{3}  \; (\bfZ_p)^2  + {\rm a.h.}
\eeq

The  mass deformation preserves the  $SO(3)\cong SU(2) $ subgroup of 
$SU(3)\times U(1)$ ${\cal R}$-symmetry. Since the  
  mass parameter is   dimensionful, the $U(1)$ symmetry is explicitly broken.  
The three dimensional 
representations 
of $SU(3)$ split  as $ {\bf 3} (  {\bf \mybar 3}) \rightarrow  {\bf 2} \oplus  {\bf 1}$ under $SU(2)$. 
  The $SO(4) \simeq SU(2)\times SU(2) $ 
${\cal R}$-symmetry, just like $\CQ=4$ supersymmetry,  is  not   harmed. 
  This means, the mass  deformed matrix model has a 
manifest  $[SU(2)]^3$ symmetry. The ten bosons and sixteen fermions 
of the matrix model, under  $[SU(2)]^3$ symmetry decompose  as 
\beq
{\bf 10} \ \longrightarrow 
 2 ({\bf 2,1,1})  \oplus  2 ({\bf 1,1,1})
\oplus  ({\bf 1,2,2}).
\eeq
\beq
{\bf 16} \ 
\longrightarrow 
({\bf 2,2,1}) \oplus ({\bf 1,2,1}) \oplus 
({\bf  2,1,2}) \oplus ({\bf 1,1, 2})
\oplus  ({\bf 1,2,1})
\oplus  ({\bf 1,1,2})\ \ 
\eqn{branching}
\eeq
As we have seen in \S\ref{sec:twistedE}, the $[SU(2)]^3$ symmetry is the 
${\cal R}$-symmetry group of the  A-twist shown in 
 \Eq{twist-1}. 
The $U(1)'_E$ twisted rotation symmetry  is not  a symmetry of the matrix model.  It  emerges in the continuum in the same way as the Lorentz symmetry emerges in the continuum limit of a lattice gauge theory. 

\subsection{Noncommutative moduli space}
The zero action configurations of the deformed matrix theory, also called the
noncommutative moduli space,  is the locus of the bosonic action: 
\beq 
S_{bosonic}|_{\cal M} =0
\eeq
This implies the vanishing of   $F$ and $d$ terms and all other 
 bosonic positive definite terms in the action:
\beq
&&F_m= \frac {\partial W(z^m)}{\partial z^m}=  0 
,  \qquad  
d= -i\sum_{m=1}^{3} [\mybar z_m, z^m]=0,  \qquad  [v_{\mu}, v_{\nu}]=0, \qquad
[v_{\mu}, z^{m}]=0 \qquad 
\eqn{full}
\eeq
where $m=1, \ldots 3$ and  $\mu=1, \ldots 4$. 
The $F$-term conditions are  
\beq
[z^m, z^n]= - m \epsilon_{mnp} z^p 
\eqn{sphere}
\eeq
The  anti-hermitian  ${\Im}(z^m)$ satisfies the  commutation relations of the $SU(2)$ algebra. 
Since the $F$-term conditions are not homogeneous under $z \rightarrow c z$
where $c \in \mathbb C$,   the mass deformed theory 
(in the chiral multiplet sector) does not possess a moduli space, rather it has a discrete
isolated set of classical minima.
The solutions of the $F$-terms also satisfy the  $d$-term condition. 
The other  conditions in \Eq{full} put certain restrictions on the form 
of $v_{\mu}$, but not $z_m$.

The  \Eq{sphere} has both  reducible and irreducible set of  solutions \cite{Vafa:1994tf}.  
For example, an irreducible embedding of $SU(2)$ algebra   into $U(N)$ 
yield a   target theory with $U(1)$ gauge group. 
In order to  construct the continuum $U(k)$ gauge theory on $S^2$, it is more convenient to 
start with   a $U(Nk)$ matrix model and expand the fluctuation around the 
rank-$N$ background solution of \Eq{sphere}.  Formally, we have 
\begin{equation} 
U(Nk) \longrightarrow \underbrace{U(N)}_{S^2 \;  \rm background} \otimes \underbrace{U(k)}_{\rm fluctuations}  
\end{equation}
The background solution for  \Eq{sphere} can explicitly be written as 
\beq 
z^p= i m J^p, \qquad {\rm where} \;\;\; [J^m, J^n]= i \epsilon_{mnp}J^p  
\eeq
where $J_p$ are generators of $SU(2)$ algebra. 
The irreducible embedding of $SU(2)$ into $U(N)$ is an  angular momentum  
\beq
j \equiv  \frac{N-1}{2}
\eqn{irrep}
\eeq
 representation. The 
eigenvalues of each  ${\Im}(z^p)$  ranges in the interval 
$ m \left[ -\frac{N-1}{2}, \ldots ,\frac{N-1}{2}\right]$. However,   
 these matrices do not commute 
with each other, and  consequently, the moduli space is noncommutative. The 
eigenvalues lies on the surface of a sphere (which is often referred 
as ``a fuzzy sphere'') in moduli space:  
 \beq 
\sum_{p=1}^{3} \; (z^{p})^{2}=  m^2 \sum_{p=1}^{3} \; (J^{p})^{2} =  m^2 \; j(j+1) \; 1_N 
= 
 \frac{m^2 (N^2-1)}{4} 1_N 
\eeq
where the last equality follows from \Eq{irrep}. The radius of this fuzzy sphere $R_{\rm fuzzy}$ in the moduli space is the    UV cut-off ($\Lambda_{\rm UV}$) of the  matrix regularization 
\beq
R_{\rm fuzzy} = \frac{mN}{2}= \Lambda_{\rm UV}, 
\eeq 
 up to lower order terms in $1/N$.   In the continuum limit, the size of this fuzzy sphere 
 diverges in an analogous manner  with  the Brillouin zone of a lattice gauge theory, whereas 
   the base space of our target theory has a fixed size determined by $m^{-1}$ , the IR scale.

In what follows, the non-commuting zero action configurations  $J^p, \; p=1,2,3$ play 
 two roles.  They generate the $S^2$ background and 
the ``hopping" (kinetic) terms in the target $U(k)$  
theory.

\subsection{Classical spectrum and  ``onion ring" Brillouin zone } 
\label{sec:spectrum}
In order to analyze  the quadratic fluctuations  of the action,   we expand the  superfields around the zero action configuration:
\beq
\eqn{background}
\bfZ_p= \langle \bfZ_p  \rangle +     \widetilde \bfZ_p =  im  \; J_p  \otimes {\bf 1}_k+  \widetilde \bfZ_p 
\eeq
where  $ \widetilde \bfZ_p $ denotes the fluctuations.    We expand 
 generic matrix field ${\bf X} $ in our action as   
\beq 
 {\bf X}= \sum_{l,m} Y_{lm} \otimes {\bf X}_{lm}  
\eqn{flucS}
\eeq
where $Y_{lm}$  are   $N\times N$  matrices (given below), associated with the angular 
momentum mode $l, m$. The fields  $ {\bf X}_{lm}$ is the $U(k)$ algebra valued   field associated with the  momentum $l, m$ in a spherical decomposition.

In  \Eq{background}, the fluctuation matrix $\widetilde \bfZ$  is a 
$GL(Nk, \mathbb C)$ valued matrix.   The zero action configuration matrices 
can be used to 
form a complete orthonormal    matrix 
basis for $GL(N, \mathbb C)$.  Since $GL(N, \mathbb C)$ 
 is  $N^2$ complex dimensional vector space, we need $N^2$ basis matrices.  
A complete orthonormal basis is generated by using the three $J_p$ matrices, and 
 by just mimicking the  spherical harmonics,  $Y_{lm}$. For example, for $GL(3,  \mathbb C)$, 
a complete orthogonal basis ${\cal B}(S^2)$ composed of nine three by three matrices are  given by 
\begin{equation}
{\cal B }(S^2) \equiv \Big\{1, J_z, J_{\pm}, J_{\pm}^2, J_z 
J_{\pm}, (3J_z^2 -1)\Big\} \sim 
\Big\{ Y_{00}, Y_{10}, Y_{1, \pm 1}, Y_{2, \pm2}, Y_{2, \pm1}, Y_{2,0} \Big\} 
\eqn{basisS}
\end{equation} 

The classical spectrum of the  fermions and bosons can be found by studying the fluctuations  around the background.  This is a straightforward calculation along the lines of analysis of 
\cite{Andrews:2006aw} and  \cite{Unsal:2005us}.  In particular, the details of the {\it classical} analysis  are  literally {\it identical} to the deconstruction of the Maldacena-N\'u\~nez compactification   on $\R^4 \times S^2$, starting with  
 $\CN=1^*$ SYM theory in $d=4$ dimensions,  and are  discussed thoroughly in  \cite{Andrews:2006aw}. Hence, this classical analysis  will not be repeated here. The interaction terms as well
work precisely as in \cite{Unsal:2005us}   and \cite{Andrews:2006aw}.  This means, at the classical level, our deformed matrix model produces the target theory on $S^2$ correctly. 
Below, we  discuss some interesting physical aspects of the matrix regularizations.

The spectrum of Grassmann even and odd modes of the matrix model, their level degeneracy and  their transformation properties under the twisted rotation group are given by 
\begin{eqnarray}
\begin{array}{|c|c|c|c|}
\hline
{\rm  Free \; Spectrum} &  {\rm Grassmann \; odd} & {\rm  Grassmann \; even }  & SO(2)_E'  
\cr
& & &\cr 
\hline
M^2_{l, m }= m^2 \; l(l+1), \,\,\, l=0, 1,2, \ldots  \qquad  &  
 4 (2l +1),    &   4(2l +1)   & {\rm scalar}
\cr 
& & &\cr
\hline 
 M^2_{l', m' }= m^2  \; (l'+\half)^2, \,\,\,  l'=\half, {\textstyle 
\frac{3}{2}}, {\textstyle 
\frac{5}{2}}, \ldots, \qquad   &  8 (2l' +1)   & 4 (2l' +1)   & {\rm spinor}
\cr
& & &\cr
\hline 
 M^2_{l, m }= m^2  \; l(l+1),  \,\,\, l=1, 2,3,  \ldots,  \qquad &
 4 (2l +1)   &   2 (2l +1)   & { \rm vector} \cr
& & & \cr 
\hline
\end{array} .
\eqn{spectrum}
\end{eqnarray}
The three types  of  the spectrum can be naturally associated 
with the truncated spectrum of spin-0, spin-$\half$  and spin-$1$ fields on  $S^2$. The 
spectral degeneracy  of the Grassmann even and odd   spin-0, spin-$\half$  and spin-$1$ 
modes is a consequence of the exact supersymmetry of the deformed matrix model.

{\bf  Brillouin zone:} The spectrum  shown in \Eq{spectrum} also  provides  a notion of the 
Brillouin zone for the matrix regularization.   In the  penultimate line of \Eq{spectrum}, define 
$l=  (l'+\half), \; l=1,2, 3 \ldots $. 
 The Brillouin
zone is composed of circular shells, like the onion rings and $l^{th}$ shell 
accommodates $(2l+1)$  states.  
The cut-off is determined by the size of the 
matrices in the matrix
regularization, $N$, and $j_{max}= \frac{N-1}{2}$ and the UV cut-off is 
$\Lambda_{\rm UV}= m \frac{N}{2}$.  As in the lattice regularization, 
wavelengths below the  length scale $\Lambda^{-1}_{\rm UV}$ are not present 
in the matrix regularized  theory.   
In the continuum limit, we take the cut-off $\Lambda_{\rm UV}$ to infinity while keeping
$m$ fixed (the inverse size of the two-sphere)  and  taking $N\rightarrow \infty$.
In the moduli space, this corresponds to taking radius of the 
fuzzy sphere to infinity, similar to the  deconstruction and
supersymmetric lattices where  the continuum limit is a trajectory out to infinity
in the moduli space.

\subsection{A-twist  and  mass deformation}
{\bf Fermions:} Clearly, the spectrum of the Grassmann odd variables shown in 
 \Eq{spectrum}  is {\bf not} what one  would naively  expect from the eigenvalue  spectrum of a Dirac operator  on  a two-sphere $S^2$.  Instead, it is a mix of 
truncated spectrum of spin-0, spin-$\half$  and spin-$1$ fields on a sphere. 
In \Eq{spectrum}, another  bizarre  feature at first glance is the  
appearance of fermion  zero modes.  However, it is well known that, 
the eigenvalue spectrum of the 
Dirac operator  on $S^2$, (and in general in any  
positively curved background)  has a gap  due to spin connection. \footnote{
More  generally, the Dirac operator on a curved background is given by  
$\Dslash =  \gamma^{a} e_{a}^{\;\; \mu} D_{\mu}$ 
 where  
$D_{\mu} = \partial_{\mu} + \omega_{\mu}$ 
is the general 
covariant derivative and $\omega_{\mu}$ is 
the spin connection. ${\mu}$  is the global coordinate index and   
$a$ is  local frame index.  The spin connection is $\omega_{\mu} = 
\omega_{\mu}^{\;\; ab} \Sigma_{ab}$ where  $\Sigma_{ab}$ are generators 
of local rotations (acting in spinor representation). It is a simple exercise 
to show that the eigenvalue spectrum of the Dirac operators has a gap.}

Of course, although  the fermionic  spectrum sounds incorrect  for a naive (no supersymmetry preserving) compactification of the ${\cal N}= (8,8)$ theory on $S^2$, it is on the other hand precisely what one expects from the  compactification of the  twisted formulation 
 discussed  in \S\ref{sec:twist1}. 
Due to  twisting, we have   global supersymmetry  on $S^2$, the  
   spectrum of fermions has 
 four fermionic zero modes, which is in exact correspondences 
  with the presence  of  four exact supersymmetries of the matrix model \Eq{deformation2}. 

{\bf Bosons:}  
The  spectrum of bosons coincides  with the 
truncation of the  A-twist \Eq{twist-1} of the supersymmetric theory, but not 
with the naive untwisted compactification. Of course, this  is  consistently tied with what we have presented for fermions in terms of twisting. 
  \footnote{
The discussion of this section can be easily generalized to target theories on $\R \times S^2$. 
For a  very interesting proposal about $\CN=4$ SYM theory on $\R \times  S^3$, see  
\cite{Ishii:2008ib}.}

\subsection{B-twist  and $\beta$-flux  deformation and target theory on $T^2$}
\label{Sec:B}
The $\beta$-flux  deformation of the superpotential is a one-parameter family of deformation given by 
\beq
\eqn{deformation3}
W( {\bfZ}^m)_{\beta} &&= \sqrt 2 {\bfZ}^1  ( e^{-i \beta/2}  \bfZ^2 \bfZ^3 - e^{+i \beta/2}  \bfZ^3 \bfZ^2 ) 
\eeq
or equivalently, 
\beq
   W( {\bfZ}^m)_{\beta}      = \frac{\sqrt 2}{3!} \epsilon_{mnp}  \, {\bfZ}^m  (\; e^{-i \Phi_{np} /2}   \bfZ^n  
\bfZ^p\; -   e^{+i \Phi_{np} /2}   \bfZ^p  
\bfZ^n\;  ) 
\eeq 
with obvious identifications. 
The deformation is respectful to $SU(2)_L \times SU(2)_R$ ${\cal R}$-symmetry of the matrix model, 
whereas it  only preserves $U(1)^3$  subgroup of  the  $SU(3)\times U(1)$ symmetry. Recall from 
\S\ref{sec:twist2} that 
the $SU(2)_L \times SU(2)_R$   is also the non-abelian   global ${\cal R}$-symmetry of the 
B-twist .

 The $\beta$ flux-deformation does not introduce a dimensionful 
parameter,  
unlike the case with mass deformation.  
However, the $\beta$-flux  deformed theory possesses 
a degenerate manifold of the ground states,   a moduli 
space, where the distance from the origin of the moduli space 
has an interpretation as an  UV cut-off,  
similar  to the supersymmetric orbifold lattices.

The zero action configuration of the $\beta$-flux deformed theory is the solution 
of  \Eq{full}. Given  the  superpotential  \Eq{deformation3}, the  $F$-term constraints reduce to   
 a slight generalization of the  't Hooft  algebra 
\beq 
z^1 z^2  = e^{i\beta} z^2 z^1, \qquad   z^2 z^3 = e^{i\beta} z^3 z^2, \qquad 
z^3 z^1 = e^{i\beta} z^1 z^3
\eqn{tek}
\eeq
Let us consider a $U(Nk)$ deformed matrix theory. Our goal is, similar to the 
mass deformed matrix model, to generate a base space and gauge theory residing on it: 
\begin{equation} 
U(Nk) \longrightarrow \underbrace{U(N)}_{T^2 \;  \rm background} \otimes \underbrace{U(k)}_{\rm gauge  \; fluctuations}  
\end{equation}
 
Such constructions at the classical level are  standard, for example, for the classical relation 
between   non-supersymmetric TEK model regularization to the non-commutative Yang-Mills theory,  see the  Refs.\cite{Azeyanagi:2008bk, Bietenholz:2006cz, Unsal:2004cf} and references therein.   The non-commutative Yang-Mills theory also possess the commutative limit for appropriate choice of deformation parameters. 
The construction of the square lattices on $T^2$ in matrix regularization is well-known. Below, we additionally point out how to construct the hexagonal 
$A_2^*$ lattice without getting into details.  Of course, the main point of this section is that  
$\beta$-flux deformation produce the B-twist in its continuum limit.

Let us choose  the  deformation matrix as 
\beq
\Phi_{np}= \left[ \begin{array}{cc|c}
     & +\frac{2 \pi}{N}  & - \frac{2 \pi}{N}  \\
     -\frac{2 \pi}{N}  &  & +\frac{2 \pi}{N}   \\
\hline     
+ \frac{2 \pi}{N} & -\frac{2 \pi}{N}  & 
       \end{array} \right]_{np}
\eeq
The solutions  of the 't Hooft  algebra is given in terms of clock and shift
matrices:
\begin{equation}
(P)_{kl}= e^{ i2 \pi k/N} \delta_{kl}, \qquad  (Q)_{kl}=  
\delta_{k+1,l},  \qquad k,l= 1, \ldots N  
\eqn{clock}
\end{equation}
We background matrices are 
\beq
\langle z^1 \rangle= c^1 P \otimes 1_k, \qquad 
\langle z^2  \rangle = c^2 Q \otimes 1_k \qquad
\langle z^3 \rangle = c^3 (PQ)^{-1} \otimes 1_k
\eeq
where $c^i\in \mathbb C$ are complex modulus parameters which are essential
in establishing a continuum  limit.  The presence
of these moduli fields is expected.  If  \Eq{tek} has a solution, due  to its 
homogeneity under  $z^m \rightarrow \alpha z^m$,  it has a continuum of solutions (unlike the   mass deformed theory \Eq{deformation}).  
The conditions $[v_{\mu},z^{m}]=0$ restrict  $v_{\mu}$ to a 
matrix proportional to identity. Since $v_{\mu}$ matrices commute with both $P$ and
$Q$,  they must  be proportional to the Casimir of the 't Hooft algebra, which
is identity. With this configuration of the matrices, the  $d$-term constraint and  $[v_{\mu}, v_{\nu}]=0$ are automatically satisfied. 
Consequently, classical  moduli space is (for $k=1$)
\beq
{\cal M}= {\mathbb R}^{4}\times  {\mathbb C}^{3} 
 \eeq
The existence of    ${\mathbb C}^{3}$  along which one can move to infinity is sufficient to 
produce a continuum $\CN=(8, 8)$ theory.  

We can map the matrix model to 
two  types of   lattices. 
The analog of the basis for $S^2$  \Eq{basisS} can now be constructed by using the clock and 
shift matrices.
  The basis matrices are  $J_{(p_1,p_2)}\sim Q^{p_1} P^{p_2}, \; p_i=1, \ldots N$ , where 
 $p_i$ gains interpretation as  momentum  in a two dimensional  Brillouin zone.
 In particular, expressing the  fluctuations of the matrix fields as 
$
\widetilde {\bfZ}= \sum_{{\bf p}} J_{\bf p} \otimes \widetilde \bfZ_{\bf p} 
$
produces the  $\CQ=4$ supersymmetry preserving  
non-commutative lattice regularization for the  $\CN=(8,8)$ target theory. 
 As usual,   the square and $A_2^*$ lattices emerges by expanding around 
the following points in the moduli space 
\beq
&& {\rm Square \; \;  lattice: }  \; \; c_1= c_2 = \frac{1}{a}, \;\; c_3=0, \\
&& {\rm Hexagonal}  \; A_2^* \; \;  {\rm lattice}: \;  c_m = \frac{1}{a}, \qquad m=1,2,3
 \eeq
 The details of this type of  calculations can be found in 
\cite{Unsal:2005us}. 

\subsection{Comments} 
There are a few points that we wish to emphasize in  this construction: 

{\bf 1)} The matrix regularization given in   \Eq{deformation3} has only $\CQ=4$ exact 
supersymmetries. These are the scalar supersymmetries of the B-twist version of the 
target theory.     Since the target theory is defined on a flat $T^2$ (or $\R^2$ in its infinite volume limit),  the other 12 (non-scalar)  supersymmetries arises accidentally in the continuum.  
Note that as it is in $d=4$ dimensional $\beta$-deformation (the Leigh-Strassler deformation 
\cite{Leigh:1995ep}), the 
remaining twelve  supersymmetries are explicitly broken, and are not symmetries of the matrix model    in any sense.
 
{\bf 2)} In matrix model  approach, there is no orbifold projection. 
  The   $N^2 k^2$  total number  of  microscopic degrees of freedom  of the $U(Nk)$ matrix model 
  transmutes into a non-commutative $U(k)$    lattice  gauge theory with  $N^2$  sites.  
  In the orbifold projection, in order to generate  a two  dimensional regular lattice, 
  one starts with $U(N^2k)$ matrix model, which has $  N^4k^2$  degrees of freedom  and projects out  by a $(Z_N)^2$ discrete symmetry:
  \begin{equation} 
U(N^2k) \underbrace{\longrightarrow}_{\rm orbifolding}  [U(k)]^{N^2}
\end{equation}
In the matrix regularization, one keeps all the degrees of freedom of the matrix model and in the latter, one removes most degrees of freedom by projections.  Orbifolding   results in an ordinary 
SYM theory on a commutative (regular) lattice with nearest neighbor interactions. 
The  matrix regularization  has both commutative and non-commutative continuum limits.

{\bf 3)} The $\beta$-deformation of the $d=0$ matrix model 
 can only produce the target theories in {\it even} dimensions, $d=2, 4$. 
  In orbifold constructions, there is no distinction between even and odd dimensional target theories. \footnote{ If one starts with matrix quantum mechanics with $d=0+1$, one can only produce Hamiltonian formulations in 
 $d=2+1$ dimensions.}

{\bf 4)} It should be noted that our  analysis of the fluctuations on both  ($S^2$ and $T^2$)
backgrounds  is classical.  
As explained, the mass deformed  action  
has many discrete, isolated minima corresponding to different background configuration of the matrix degrees of freedom and the $\beta$ deformed theory has a classical moduli space.   
As we have argued, not all these background zero action configurations lead to a regularized 
field theory on   $S^2$ and $T^2$.  In our classical  analysis,  we {\it choose} to expand 
around a particular minima \Eq{background}. It is in principle  possible that the statistical 
fluctuations can take the equilibrium state we expand around  to another one
 which does not have ``an emergent space" interpretation and hence spoil the whole picture.

Indeed, Refs.~\cite{Bietenholz:2006cz, Azeyanagi:2008bk}
  recently showed  a non-perturbative instability in related bosonic matrix models.     The TEK model, which produces the $d=4$ dimensional  non-commutative YM theory in its {\it classical} continuum limit, fails to be stable non-perturbatively. 
 Ref.~\cite{Azeyanagi:2008bk}  also shows that in supersymmetric matrix models, the non-commutative  background is stable (even if supersymmetry is broken softly).  In this sense, the manifestly supersymmetric matrix model regularizations of the supersymmetric theories should be  producing a stable background.  
According to the criteria of Ref.~\cite{Azeyanagi:2008bk}, both of our supersymmetric target theories as well as the supersymmetric deformed  matrix models of Refs.~\cite{Nishimura:2003tf,Unsal:2004cf,Unsal:2005us} are safe. 
For the detailed discussion, we refer the reader to Ref.\cite{Azeyanagi:2008bk}.

\section{A new class of supersymmetric gauge theories: $\CN = \fourth$  SYM} 
In this section, motivated by  the recent advances in supersymmetric lattice 
constructions and using ideas from the topological  field theories,  
we  define a new class of supersymmetric theories with  $\CN = \fourth$  supersymmetry on $\R^4$.  This construction will be used to address certain  questions about exact lattice supersymmetry,  although   it may   have a wider class of applications. 

The approach described in what follows can be applied to   extended supersymmetric gauge theories in  various dimensions. We will describe it in $d=4$  dimensions, starting with  $\CN = 4$  SYM theory. 

The  proposal is as follows: First, we  twist the  $\CN = 4$  SYM theory formulated on $M=\R^4$.  Then, we  deform the action on  $\R^4$ such that only one out of sixteen supersymmetries is preserved exactly.

Recall that the  twisted theories on flat space-times such as   $M= \R^4,  \; T^4$   are  simply a 
 rewriting of the original theories in terms of representation of the new Lorentz group.
 The twisted theory on $\R^4$ preserves the  same set of supersymmetries as in the 
 original theory, and the twist can be undone. 
A well-known way to preserve only the scalar sub-set of supersymmetry is to carry the theory 
into curved space. This is in essence same as declaring the scalar supersymmetry as 
some type  of   BRST operator.  Here, we will not do so. Instead, we will simply deform the twisted action on $\R^4$ in such a way that only  $\CN = \fourth$ is respected.\footnote{This proposal is different from  Seiberg's 
$\CN= \half$ construction. See \S.\ref{seiberg}.}

\subsection{Twist the algebra,  deform the action: From $\CN = 4$   to   $\CN = \fourth$  on $\R^4$  } 
\label{td}
{\it Twisting}
\\
The $\CN = 4 $ theory on $\R^4$ can be obtained as the dimensional reduction of the 
$\CN=1 $ gauge theory on $\R^{10}$ down to $\R^4$. The ten dimensional theory possess an 
$SO(10)$ Euclidean Lorentz rotation group. Upon reduction,  the $SO(10)$ group decomposes
 into 
 \beq
SO(10) \longrightarrow 
\left( \begin{array}{c|c}
     SO(4) & \\
\hline     
 & SO(6)
       \end{array} \right)
\eeq
where $SO(4) \sim SU(2)_L \times SU(2)_R$  is the four dimensional Lorentz symmetry action 
on $\R^4$ and $SO(6)_{\cal R} \sim SU(4)_{\cal R}$ is the internal ${\cal R}$-symmetry group. \footnote{We do not  distinguish the orthogonal groups from the spin groups. Whatever is implied will be clear from the context. }

The ${\bf 16}$ dimensional positive chirality spinor of $SO(10)$ and the  sixteen supercharges 
decompose as 
\begin{equation}
  Q_{\alpha, I} \oplus  \mybar Q_{\dot \alpha, I} \sim    ({\bf 2,1,4})   \oplus  ({\bf 1,2,\bar 4})  \in 
   SU(2)_L \times SU(2)_R \times SU(4)_{\cal R}
   \eqn{super}
  \end{equation}
The twisting procedure is a choice of an  $[SU(2) \times SU(2)]'$ embedding into  $SU(2) \times  SU(2) \times SU(4)_{\cal R}$. There are three inequivalent twists of $\CN=4$ SYM 
\cite{Vafa:1994tf}, only 
one of which emerges naturally from supersymmetric lattices,  and the two others do not.  The reasons is discussed in  detail in \S\ref{sec:SL}.  

The  twist which arises naturally in supersymmetric lattices maps all the supercharges (and fermions) into integer spin representation. 
This correspond to the Dirac-K\"ahler decomposition of multiple-spinors as often used in 
lattice gauge theory. This  twist   arises naturally on $A_4^*$  or hyper-cubic lattice
definition of the $\CN=4$ SYM theory.    In order to distinguish the twists which admit a lattice implementation  \cite{Marcus:1995mq} and the ones which do not \cite{Vafa:1994tf}, it seems convenient to address the first class 
as supersymmetric lattice twists (SL-twists). 

In what follows, let us choose an SL-twist.  It is most easily described by the 
  decomposition of ${\bf 4}$ of $SU(4)_{\cal R}$ into  $({\bf 2,1})   \oplus  ({\bf 1,2})$ and 
by the diagonal embedding of the twisted Lorentz group,   
\begin{equation}
 [SU(2) \times  SU(2)]'  \subset {\rm Diag} \Big( [SU(2) \times  SU(2)]_{\rm Lorentz} \times   [SU(2) \times  SU(2)]_{\cal R} \Big) \; .
\end{equation} 
The spinors (and supercharges) decompose into $p$-form integer spins:
\begin{equation}
  Q_{\alpha, I} \oplus  \mybar Q_{\dot \alpha, I} \longrightarrow  Q^{(0)} \oplus      
  Q^{(1)}  \oplus  Q^{(2)}  \oplus  Q^{(3)}  \oplus  Q^{(4)} 
\end{equation}
The twisted supersymmetry algebra in four dimensions 
has one or two  nilpotent scalar subalgeras, a particularly useful one being 
\begin{equation}
  (Q^{(0)})^2 \equiv Q^2 = 0
  \end{equation}
which does not care about the background spacetime,  and is a charge (which is defined
globally) even if the background space is curved or discrete. The higher form supersymmetries, 
$Q^{(1)}$ for example, cannot be globally  defined on a curved space, because of the absence of the covariantly constant four vectors on four manifolds. $Q^{(1)}$ cannot be globally defined on a lattice either, since the  anti-commutator   $\{ \eta Q,      
  \eta_{\mu} Q^{{\mu}} \}  \; \cdot  \sim \eta \eta_{\mu} P^{\mu} \; \cdot $ is an infinitesimal translation, and there are no infinitesimal translation on the lattice. 
This tells us that the exact global supersymmetry that can be achieved on  lattice and on curved spaces are  necessarily the  scalar subalgebra.   

We label the fermionic  matter content of the twisted theory 
as $p$-form Grassmann variables  
$( \lambda,  \psi^{\mu},  \xi_{\mu \nu},  \xi^{\mu \nu \rho},  \psi_{\mu \nu \rho \sigma})$. The 
bosonic content is $( z^{\mu} ,  \mybar z_{\mu} ,   z_{\mu \nu \rho \sigma} 
\mybar z^{\mu \nu \rho \sigma} )$ where $z^{\mu}= (S^{\mu} + i V^{\mu})/ \sqrt 2  $ is a complexified gauge field  which is the linear combination of the gauge boson and four scalars
of the original theory.  The other scalars are the fully anti-symmetric  $ z_{\mu \nu \rho \sigma} $
and its conjugate. We  also need  complex gauge covariant derivative 
$\CD^{\mu}\, \cdot  = \partial^{\mu} \cdot + \sqrt 2 [ z^{\mu},   \, \cdot \,], $ and associated 
two-form field strength $\CF^{\mu \nu}= -i [\CD^\mu, \CD^\nu]$.  

The  (off-shell) action of the 0-form supercharge $Q$ is 
 \begin{eqnarray}
&&Q \lambda =  -id,   \qquad    Q d = 0 \cr 
&& Q z^{\mu} =   \sqrt 2 \, \, \psi^{\mu}, \qquad  Q \psi^{\mu}=0 \cr
&& Q \mybar z_{\mu} = 0  \cr
&&Q \xi_{\mu\nu} = -i \mybar \CF_{\mu \nu} \cr
&&Q \xi^{\nu \rho \sigma} = \sqrt 2 \, {\mybar \CD}_{\mu}  
\mybar z^{\mu \nu \rho \sigma}  \cr
&&Q z_{\mu \nu \rho \sigma} = \sqrt{2} \psi_{\mu \nu \rho \sigma}, \qquad 
Q \psi_{\mu \nu \rho \sigma} = 0    \cr 
&&Q \mybar z^{\mu \nu \rho \sigma} =0  
\eqn{QAoffshell}
\end{eqnarray}
where $d$ is an auxiliary field introduced for the  off-shell completion of 
the scalar  supersymmetry subalgebra.  

{\it Deform}
\\
The  $\CN=4$ SYM lagrangian on $\R^4$ can be expressed in a way to make only $\CN=\fourth$
 manifest.  Obviously, with the spinor supercharges, the minimal amount of supersymmetry that 
 we can have in a supersymmetric theory is  $\CN=1$. This constraint can be circumvented upon having a spin-0 scalar supercharge.

The twisted  Lagrangian on $\R^4$  may be written as a sum of $Q$-exact and  $Q$-closed 
terms:
\beq
 \CL = && \CL_{exact} +\CL_{closed} =  \CL_1 +  \CL_2 +   \CL_3 =  
Q {\widetilde \CL_{exact}} +  \CL_{closed},
\eqn{LT41} 
\eeq 
where 
 $\widetilde \CL_{exact}= {\widetilde \CL}_{e,1} + 
{\widetilde \CL}_{e,2} $ is given by  
\beq
&& {\widetilde \CL}_{e,1} =    \frac{1}{g^2}\Tr  \Big( \lambda ( \half id + \half 
[ \mybar {\cal D}_{\mu}, {\cal D}^{\mu} ] +
\textstyle{\frac{1}{24}}[\mybar z^{\mu\nu \rho \sigma}
, z_{\mu\nu \rho \sigma}] ) \Big) \cr
&& {\widetilde \CL}_{e,2}=   \frac{1}{g^2}
  \Tr  \Big( 
\fourthi \xi_{\mu\nu} \CF^{\mu\nu} +  \textstyle{\frac{1}{12 \sqrt2}}
\xi^{\nu \rho \sigma} {\cal D}^{\mu} z_{\mu \nu \rho \sigma} 
\Big) 
\eeq
and   $\CL_{closed}$ is given by 
\beq
 \CL_{closed}= \CL_3 =  \frac{1}{g^2} \Tr 
\half \xi_{\mu\nu }  \mybar {\cal D}_{\rho} \xi^{\mu \nu \rho} +  
\textstyle{\frac{\sqrt2}{8}} \;  \xi_{\mu\nu } [\mybar 
z^{\mu \nu \rho \sigma}, \xi_{\rho \sigma}] \qquad
\eeq
and $g$ is coupling constant. 
By using the transformation properties  of fields and 
the equation of motion for the auxiliary field $d$,  
we obtain the Lagrangian expressed in terms of propagating degrees of 
freedom:
\beq 
 &&  \CL_1 =     \frac{1}{g^2}\Tr  \Big( \half (  \half [\mybar {\cal D}_{\mu} , 
 {\cal D}^{\mu}] + \textstyle{\frac{1}{24}}[\mybar z^{\mu\nu \rho \sigma}
, z_{\mu\nu \rho \sigma}])^2 +   
\lambda ( \mybar {\cal D}_{\mu} \psi^{\mu} +   
\textstyle{\frac{1}{24}}[\mybar z^{\mu\nu \rho \sigma}
, \psi_{\mu\nu \rho \sigma}]) \Big)  \cr 
&&  \CL_2 =    \frac{1}{g^2} \Tr  \Big(  \fourth \mybar {\CF}_{\mu\nu} {\CF}^{\mu\nu}    
+   \xi_{\mu\nu}  {\cal D}^{\mu} \psi^{\nu} 
+ {\textstyle \frac{1}{12}} |{\cal D}^{\mu} z_{\mu \nu \rho \sigma}|^2
+  {\textstyle \frac{1}{12}} 
 \xi^{\nu \rho \sigma} {\cal D}^{\mu} \psi_{\mu \nu \rho \sigma} 
+  {\textstyle \frac{1}{6 \sqrt 2 }}
 \xi^{\nu \rho \sigma} [\psi^{\mu}, z_{\mu \nu \rho \sigma}]
\Big)   \cr 
&&  \CL_3 =  \frac{1}{g^2}  \Tr \Big(
\half \xi_{\mu\nu }  \mybar {\cal D}_{\rho} \xi^{\mu \nu \rho} +  
\textstyle{\frac{\sqrt2}{8}} \;  \xi_{\mu\nu } [\mybar 
z^{\mu \nu \rho \sigma}, \xi_{\rho \sigma}] 
\Big) \; .
\eqn{LT4}
\eeq

The $Q$-invariance of the  $\CL_{exact}$ is obvious and  
 follows from supersymmetry algebra $Q^2=0$.
To show the invariance of $Q$-closed term  
requires the  use of the Bianchi  identity.   
The Lagrangian \Eq{LT41} possesses  a manifest $\CN=\fourth$ supersymmetry,   $SO(4)'$   
 twisted Lorentz symmetry, and  $U(1)_{\cal R}$ 
${\cal R}$-symmetry.  

This form of the $\CN=4$ Lagrangian as well as its  generalizations by fermionic 
symmetry  
\beq
Q(u,v)  \sim u Q^{(0)} +  v (*Q^{(4)}), \qquad u, v \in {\mathbb C}  
\eqn{uv}
\eeq 
  where $*$ is Hodge-dual 
had multiple useful  applications during the recent years.  The  fermionic 
symmetry satisfies 
\begin{equation}
[Q(u,v)]^2 \;  \cdot \sim uv \{ Q^{(0)} ,  (*Q^{(4)})\} \; \cdot \sim uv \delta_{\bar z} \; \cdot
\end{equation}
 where 
$\bar z = \frac{1}{4!}\epsilon_{\mu \nu\rho \sigma} {\bar z}^{\mu \nu\rho \sigma}$ and 
$\delta_{\bar z} $ is field dependent infinitesimal gauge transformation. 
This means, modulo gauge transformations, $[Q(u,v)]^2 =0$.
 Such 
generalizations of this twist at special values of the complex parameters 
were used in studying  dualities in $\CN=4$ SYM   \cite{Kapustin:2006pk} and in the comparison of 
 $A_4^*$  supersymmetric orbifold lattices  \cite{Kaplan:2005ta} 
and geometric formulation \cite{Catterall:2005fd} in 
Refs.\cite{Unsal:2006qp, Catterall:2007kn}.
These two supersymmetric lattice formulations correspond to $(u, v)\sim(1,0)$ and $(u, v)\sim(1,1)$, respectively.  
The 
 \Eq{LT41} is also the continuum limit of the  supersymmetric matrix model regularization 
of  $\CN=4$ SYM theory \cite{Unsal:2005us}. \footnote{One other interesting applications may be to instantons in $\CN=4$ theory and its dimensional reductions.
 The fixed points of the $Q(u,v)$-action
in the supersymmetry transformation gives complexified instanton equations such as 
$u {\cal \mybar F}^{(2)}  + v *{\cal F}^{(2)} =0$, or in components, 
$u {\cal \mybar F}_{\mu \nu}+ v  \half \epsilon_{\mu \nu \rho \sigma} {\cal F}^{\rho \sigma}=0$ \cite{Unsal:2006qp,Kapustin:2006pk}.}

Let us now consider  a  deformation of the action \Eq{LT41} into 
\beq
 \CL^{\CN=\fourth} =  d_1 \CL_1 +  d_2 \CL_2 + d_3  \CL_3   
\eqn{LT42} 
\eeq 
where $d_i$ are real parameters.  If the deformation parameters are equal, this is the original 
Lagrangian with a rescaled coupling constant $\frac{1}{g^2} \rightarrow \frac{d}{g^2}$. 

For unequal deformation parameters,  \Eq{LT42} is a theory with  $\CN=\fourth$ supersymmetry, as can be shown by explicit calculation on any four manifold $M$. 
 If $M$ is flat, such as  $T^4$ or $\R^4$,  for generic values of the deformation parameters, 
 the twisting cannot be undone. Hence, this is truly a theory with $\CQ=1$ (or  $\CN=\fourth$) supersymmetry even on flat spacetime.  In this sense, it is different from the 
 topological twists, which on the flat spacetime is a rewriting of the original gauge theory. 
 
 The  \Eq{LT42} seems like a BRST gauge fixing of 
  a complexified gauge invariant gauge theory. I have attempted to construct  such a BRST gauge fixing   and failed.
   Currently, the physical interpretation of the deformed Lagrangian 
  is also unclear.   \footnote{
  The ambiguity   of a BRST-like  interpretation, despite the BRST-like role of the spin-0 supercharge $Q$,   is not special to the above construction. Indeed, in the original 
  construction  of the   relativistic topological field theories from scratch, related 
   interpretational question appeared in Ref.\cite{Witten:1988ze}. A definitive answer along these lines is still lacking. }
Despite these subtleties, this Lagrangian will be useful in addressing some questions about lattice supersymmetry.

\section{Matrix model regularization for $\CN=4$ SYM in $d=4$} 
\label{matrixm}
As the $\CQ=16$ matrix model can be written in terms of $\CQ=4$ superfields, which is suitable 
for $\CQ=4$ supersymmetry preserving deformations, it can also be written in terms of 
$\CQ=1$ superfields. 
A generalization of the  $\beta$-flux  deformation  to generate four dimensional 
target theories may be used to create either an
hyper-cubic lattice  or more symmetrical $A_4^*$ lattice. 
What follows is a   concise reformulation of 
   earlier work \cite{Unsal:2005us}.

The deformed matrix model  action with $\CQ=1$ exact supersymmetry is given by  
\begin{eqnarray}
S^{\rm DMM} &=& \frac{\Tr}{g^2}  \bigg[ \int \;  d \theta \left( 
-\frac{1}{2} {\bf \Lambda} {\partial}_{\theta} {\bf \Lambda }  
- {\bf\Lambda }[ \mybar z_{m} , 
{\bf Z}_{m} ] + 
\frac{ 1}{2}
{\bf \Xi}_{mn }{\bf E}^{mn}
\right) \cr \bigg.
&+& \bigg.
  \frac{\sqrt 2 }{8}  \epsilon^{mnpqr} {\bf \Xi}_{mn} 
( e^{-i( \Phi_{pq} + \Phi_{pr}) /2} \mybar z_{p}  {\bf \Xi}_{qr}  -  e^{+i( \Phi_{pq} + \Phi_{pr}) /2}
{\bf \Xi}_{qr}  \mybar z_{p } )
\bigg] \qquad \qquad
\eqn{matrixN=4}
\end{eqnarray}
where the $\CQ=1$ supersymmetric matrix multiples are 
    \begin{equation}
\begin{aligned}
& {\bf \Lambda} = \lambda  -i\theta  d  \ ,\\
& {\bfz}^{m} = z^{m}  + \sqrt{2}\,\theta \,
  \psi^{m} , \qquad  \mybar z_{m},  \qquad m=1, \ldots,5 \\ 
 &  {\bf \Xi}_{mn}= \xi_{mn} -  
2\theta\,\, \mybar E_{mn} \, . 
\end{aligned}
\eqn{superfields}
\end{equation}
The  $\mybar z_{m}$ is supersymmetry  singlet, and hence a multiplet on its own right. 
The fermi multiplet  ${\bf \Xi}_{mn}$ is anti-symmetric in its indices.  
 The holomorphic ${\bf E}^{mn}$ functions        are the analogs of the derivative of the 
 superpotential $\epsilon^{mnp}\frac{\partial W( {\bf Z})}{\partial \bfZ^p}$  and   given by
\beq 
&{\bf E}_{mn} ( {\bf Z})&=  e^{-i\Phi_{mn}/2} {\bf Z}^{m} {\bf Z}^{n} -  e^{+i\Phi_{mn}/2}
{\bf Z}^{n} 
 {\bf Z}^{m} , \cr \cr
& { \mybar E}_{mn} ( {\mybar z})& =  e^{-i\Phi_{mn}/2} {\mybar  z}_{m} {\mybar z}_{n} -  e^{+i\Phi_{mn}/2}
{\mybar  z}_{n} 
 {\mybar  z}_{m}  \; .
 \eqn{Efunc}
\eeq
The \Eq{matrixN=4} is the $\CQ=1$ supersymmetry preserving 
deformed matrix model formulation of the target $\CN=4$ SYM theory. 

A convenient choice  for the  gauge group  of the deformed matrix model is $U(N^2 k)$ 
and a choice of flux   matrix  with a   commutative  continuum limit is    
\beq
[\Phi_{mn}]= \left[ \begin{array}{cc|cc|c}
     & +\frac{2 \pi}{N}  & &  &- \frac{2 \pi}{N}  \\
     -\frac{2 \pi}{N}  & & &  &  +\frac{2 \pi}{N}   \\
\hline       &&& +\frac{2 \pi}{N} & -\frac{2 \pi}{N}  \\
      &&     -\frac{2 \pi}{N}  & &     +\frac{2 \pi}{N}  \\     
      \hline  
         +\frac{2 \pi}{N}  &  -  \frac{2 \pi}{N} &   +\frac{2 \pi}{N}  & - \frac{2 \pi}{N}     &   
 \end{array} \right]
\eeq
With this choice of the flux matrix,  the background solution is given in  \cite{Unsal:2005us}. 
Splitting the background and fluctuations of the matrix field in \Eq{matrixN=4} and following 
similar steps  in  \cite{Unsal:2005us}, we obtain the corresponding  lattice gauge theory action:
\beq
&& S = \frac{1}{g^2}\Tr \sum_{\bfn}  \int   d \theta \left( 
-\frac{1}{2} {\bf \Lambda}(\bfn) \star {\partial}_{\theta} {\bf \Lambda}(\bfn)  
- {\bf\Lambda} (\bfn) \star \Bigl[
\mybar z_m(\bfn - \bfmu_m) \star 
{\bf Z}^m(\bfn - \bfmu_m)
 - {\bf Z}^m(\bfn) \star  \mybar z_m(\bfn) \Bigr]\right. \cr &&\qquad\qquad\qquad\qquad
 + \left.
\frac{ 1}{2}
{\bf \Xi}_{mn}(\bfn) \star \Bigl[{\bf Z}^m(\bfn) \star {\bf Z}^n (\bfn + \bfmu_m)- {\bf
 Z}^n(\bfn) \star
 {\bf Z}^m(\bfn + \bfmu_n)\Bigr]
\right) \cr
&&
 +
  \frac{\sqrt 2 }{8}  \epsilon^{mnpqr}{\bf \Xi}_{mn} (\bfn)\star
 \Bigl[\mybar z_p (\bfn-  \bfmu_p)\star  {\bf \Xi}_{qr}( \bfn +  \bfmu_m + 
\bfmu_n )
-{\bf \Xi}_{qr}(\bfn -  \bfmu_q - \bfmu_r) \star  \mybar z_p(\bfn +  
\bfmu_m + \bfmu_n) \Bigr]
\cr&&
\eqn{d4latss}
\eeq
where $\bfn$ is site index, $(\bfmu_m)_{\nu}= \delta_{m \nu} - \delta_{m5}$  for   
$m=1, \ldots , 5, \; \nu=1, \ldots, 4$. 
This is precisely the $\CQ=1$ supersymmetric lattice action of Ref.\cite{Kaplan:2005ta}  
with identical  notation therein, however, with a modified (non-local) product of lattice superfields. 
The exact $\CQ=1$ supersymmetry of the deformed model is same as the  exact lattice supersymmetry of the lattice formulation. 
The $\star$-product is encoded into a kernel $K( {\bf j -n, \; k-n})$  
\beq
\Psi_{1}({\bfn}) \star  \Psi_{2}({\bfn}) =  &&
\sum_{{\bf j}, {\bf k}}
\Psi_{1}({\bf j}) \;K( {\bf j -n, \; k-n})    \Psi_{2}({\bf k} ) \;\; \cr
\equiv&& 
\sum_{{\bf j}, {\bf k}}
\Psi_{1}({\bf j})        \left(   \frac{1}{L^4}     e^{-\frac{4 \pi i} {L^2 \theta'} \;  ({\bf j}- \bfn) \wedge    ({\bf k}- \bfn) }  \right)
 \Psi_{2}({\bf k} ) \;\;
\eeq
In this formula. 
$\theta'= 2/ N$ is a dimensionless non-commutativity parameter on the lattice, and $\wedge$ is the usual skew-product.

The \Eq{d4latss} is a   $U(k)$ lattice gauge theory on a $N^4$ lattice.   The  hyper-cubic  lattice examined in  \cite{Unsal:2005us} and the 
$A_4^*$  lattice are special points in its moduli space.  
\beq
&& {\rm Hyper-cubic  \; \;  lattice: }  \; \; c_{\mu}=  \frac{1}{a},  \; c_5=0, \qquad \mu=1, \ldots 4 \\
&&   A_4^* \; \;  {\rm lattice}: \;  c_m = \frac{1}{a}, \qquad  m=1, \ldots, 5
 \eeq
The deformed matrix model possesses   a continuum limit  which is local (or commutative). This  may be reached as 
\beq
L= N a ={\rm fixed},  \qquad N \rightarrow \infty, \;  a \rightarrow 0,    
\eeq
where we keep the size of the torus fixed. The non-commutativity parameter, in dimensionful units, is equal to 
\beq 
\Theta = \frac{N^2 a^2 \theta'}{4\pi}
\eeq 
The length scale associated with the non-locality of the $\star$-product is, 
\beq
\ell_{\star} \sim \sqrt \Theta  \sim N a \sqrt{\theta'} \sim \sqrt N a, 
\eeq
This means, in the continuum, the non-commutativity  scale tends to zero  relative to the size of the box.  For our choice of parameters, we have  
\beq
\frac{\ell_{\star}}{L}  \sim  \sqrt{\theta'} \sim \frac{1}{\sqrt N} \rightarrow 0 \; .
\eeq
By tuning $\theta'$ to be $O(1)$ in $N$ 
counting, we may also achieve a non-commutative $\CN=4$ SYM theory on $T^4$ or $\R^4$ as in the supersymmetric examples of Refs.\cite{Nishimura:2003tf,Unsal:2004cf}.  Unlike  the 
 TEK  matrix models which are recently shown to have an instability  
 \cite{Azeyanagi:2008bk, Bietenholz:2006cz},  the deformed matrix model shown in \Eq{matrixN=4} with appropriate choice of flux yields  a non-perturbatively stable $d=4$ dimensional non-commutative gauge theory according to the criteria of Ref. \cite{Azeyanagi:2008bk}. 

\subsubsection{Commutative versus non-commutative theories  and supersymmetry}
\label{seiberg}
We wish to   make  the relation between the $A_4^*$ formulation of Ref.\cite{Kaplan:2005ta}  and non-commutative lattice formulation given in  \Eq{d4latss} (at arbitrary $\theta'$) more precise. 
 First, let us consider  a supersymmetric gauge theory on $\R^4$, in continuum. 
    If we  change the structure of space such that the Grassmann 
even coordinates  $x_{\mu}$ are of  non-commutative type, and perform no manipulation about the  anti-commuting Grassmann coordinates, 
\beq 
[x_{\mu}, x_{\nu}] = i \Theta_{\mu\nu}, \qquad  \{\theta_{\alpha}, \theta_\beta\} =0
\eeq 
the resulting theory is on a non-commuting space, with  anti-commuting spinor coordinates. This manipulation does not alter 
the structure of the Grassmann odd-space and  we can define non-commutative versions of all supersymmetric gauge theories without upsetting the supersymmetry.

As an alternative to the above description, Seiberg proposed  a notion of non-anti-commuting spinor coordinates.   Instead of being anti-commuting,  the spinor coordinates satisfy a Clifford algebra\cite{Seiberg:2003yz}.  
 The consistency demands that the Grassmann even space coordinates 
 must be non-commuting as well, 
\beq 
  \{\theta_{\alpha}, \theta_\beta\} =C_{\alpha \beta} \Longrightarrow [x_{\mu}, x_{\nu}] = i \Theta_{\mu\nu}
  \eqn{sspinor}
\eeq 
where the latter is a consequence of the first. 
Ref.\cite{Seiberg:2003yz}  showed that the deformation of the Grassmann odd-space  is consistent with half of the supersymmetry and  termed this structure as $\CN=\half$ supersymmetry. 

  We do  {\bf not} introduce any deformation to the anti-commutativity in the Grassmann-odd space.  Hence, in our case, whatever structure exists in the Grassmann odd space remains intact  
 as we pass from commutative to  non-commutative  space backgrounds. Thus, Ref.\cite{Seiberg:2003yz}'s  proposal of getting an  $\CN=\half$ theory and our proposal of obtaining $\CN=\fourth$  theory are conceptually distinct.  In our case, we deform the 
 twisted-action such that only $\CN=\fourth$ remains as a symmetry of the theory. 
Moreover, by Morita equivalence, the theory on the non-commutative space is equivalent to a field theory on an ordinary space, where ordinary product is replaced by the  non-local $\star$-product of  fields. The Morita equivalence of the supersymmetric theories on $\R^d$ can also be extended into supersymmetric lattice  theories \cite{Unsal:2004cf}. 

This implies, we could  reach  the  $A_4^*$ lattice formulation of Ref.\cite{Kaplan:2005ta} 
by just  turning the $\star$-product in \Eq{d4latss} into an ordinary product, and this is indeed true. In both case, 
   the fermionic (scalar) coordinates  satisfies 
$\{ \theta, \theta\} =0$ and the amount of supersymmetry in these two formulations are equal. 
 We will benefit from this  simple observations   in one of the two  discussions of global supersymmetries in the   link approach.

\subsection{Matrix model regularization for $\CN=\fourth$ SYM in $d=4$}  
In \S\ref{td},  we introduced  a  $\CN=\fourth$ SYM theory on flat $T^4$ (and  $\R^4$) 
by deforming a twisted form of the action. The lagrangian  $\CL^{\CN=\fourth}$
of the target theory is given in \Eq{LT42}.  The main point of this deformation is
the fact that one {\bf cannot} undo the twist and recover the $\CN=4$ theory on $\R^4$, just like 
the twisted gauge theory on  $S^2$.  In deformed-twisted theories with only scalar supersymmetries, 
  we can indeed have a formulation in which both the matrix and lattice regularization and their continuum limits respect the same scalar sub-algebra, $Q^2=0$.     But as we will discuss in \S\ref{link},    the same is not true for the whole supersymmetry algebra. 
  
Here, we give a matrix model  for the theory given in 
\Eq{LT42}.  The action is 
\begin{eqnarray}
S^{\rm DD} &=& \frac{\Tr}{g^2}  \bigg[ \int \;  d \theta \; \;  d_1 \left( 
-\frac{1}{2} {\bf \Lambda} {\partial}_{\theta} {\bf \Lambda }  
- {\bf\Lambda }[ \mybar z_{m} , 
{\bf Z}_{m} ] \right) + 
d_2 \left( \frac{ 1}{2}
{\bf \Xi}_{mn }{\bf E}_{mn}
\right) \cr \bigg.
&+& \bigg. d_3 
  \frac{\sqrt 2 }{8}  \epsilon^{mnpqr} {\bf \Xi}_{mn} 
( e^{-i( \Phi_{pq} + \Phi_{pr}) /2} \mybar z_{p}  {\bf \Xi}_{qr}  -  e^{+i( \Phi_{pq} + \Phi_{pr}) /2}
{\bf \Xi}_{qr}  \mybar z_{p } )
\bigg] \qquad \qquad, 
\eqn{DD}
\end{eqnarray}
a  $\CQ=1$ preserving  doubly-deformed  matrix model.  Note that  applying  
the same $(d_1, d_2, d_3)$ deformation to the supersymmetric $A_4^*$ lattice construction of 
Ref.\cite{Kaplan:2005ta} produce a lattice regularization for \Eq{LT42}.  
The classical continuum limit of 
\Eq{DD} is the   $\CN=\fourth$ SYM theory. As stated earlier, the twist of  
 $\CN=\fourth$ cannot be undone due to the $d_i$ deformation. 
 The exact  supersymmetry in the matrix and lattice regularization is the scalar supercharge 
 of the twisted $\CN=\fourth$  theory, with continuum Lagrangian \Eq{LT42}.

\section{Link approach and  global supersymmetry }
\label{link}
Link approach is a lattice proposal  for the  supersymmetric gauge theories. 
According to the interpretation of  Refs.\cite{D'Adda:2004jb,D'Adda:2004ia, D'Adda:2005zk,D'Adda:2007ax,  Nagata:2007mz} and   on a matrix model  formulation in \cite{Arianos:2007nv,  Arianos:2008ai}, this formulation  
preserve the whole supersymmetry of the target theory on the lattice.  
\footnote{Also see  \cite{Nagata:2008zz} for application of link approach to the  Chern-Simons  gauge theory where part of the supersymmetry is preserved, and \cite{Nagata:2008xk} for an attempt 
to understand the quantum continuum limit. }
  More precisely, it is claimed that, 
all the supersymmetries of the target supersymmetric gauge theory can be preserved exactly 
on the lattice by modifying the Leibniz rule on the lattice.   

Recently, Ref.\cite{Damgaard:2007eh} unambiguously showed that 
the link approach and orbifold approach are indeed equivalent. Here, following \cite{Damgaard:2007eh},   we classify  the link approach and orbifold approach lattices in two category:
\begin{itemize}
{\item Link(1)  and Orbifold(1):
 Fermions associated with   sites, links, faces, etc. }
{\item Link(2) and  Orbifold(2):   
 All the fermions are associated with links. }
 \end{itemize}
According to the criteria of Ref.~\cite{Kaplan:2002wv} (item (iv) in  \S3), the theories obtained by orbifolding have
as many supersymmetries as the number of fermions on the sites. (These are the fermions with zero ${\bf r}$-charge in the nomenclature of Ref.\cite{Kaplan:2002wv}).   In this respect, 
 Ref.~\cite{Kaplan:2002wv} would say, the link(1)/orbifold(1)  has few supersymmetries and 
 link(2)/orbifold(2)  has  {\it  none}.    The claim of Refs.\cite{D'Adda:2004jb,D'Adda:2004ia, D'Adda:2005zk,D'Adda:2007ax} 
is  that with a  modified  Leibniz rule on the lattice,  one can devise a notion of ``link-supercharge". According to this modified criteria,  both classes above can be declared fully supersymmetric. Here, we wish to  question the latter claim. \footnote{As explained in 
\S\ref{sec:intro},  certain criticism  was raised in  literature \cite{Bruckmann:2006ub, Bruckmann:2006kb,Damgaard:2007eh}.  
 These discussions usually shape around the modified 
Leibniz rule, and the    modified  ``supersymmetry algebra" on the lattice  \cite{Bruckmann:2006ub, Bruckmann:2006kb,Damgaard:2007eh}. Here, we wish to avoid the  technicalities 
  about the modified Leibniz rule altogether, and give a direct  proof which shows that 
  the whole supersymmetry algebra cannot be preserved on the lattice. }

Recall that  in $d=4$ dimensions, the   $\beta$ deformation 
 and  a mass deformation  of superpotential reduce the $\CN=4$  SYM down to $\CN=1$ 
 \cite{Leigh:1995ep,Vafa:1994tf}. 
  There is no ambiguity in the amount of global supersymmetry here, 
  because  the other twelve supersymmetries are explicitly spoiled by the deformation. This can 
  be shown by explicit computation. 
We can dimensionally reduce these theories down to $d=0$ dimensional matrix models, and the amount of exact global supersymmetry is unaltered  by this reduction.  These are the matrix models 
studied in  \S\ref{SusySphere} and \S\ref{Sec:B}. We can construct a  $\CQ=1$ supersymmetry  preserving matrix model deformation of the  $\CQ=16$  matrix model  too
\cite{Unsal:2005us}. This is just a simple generalization of the   Leigh-Strassler $\beta$ deformation  \cite{Leigh:1995ep}.    

The $\CQ=1$  $\beta$-deformed matrix models are equivalent to the non-commutative hyper-cubic \cite{Unsal:2005us}   and  $A_4^*$ formulation. 
The same is also valid for $\CQ=4$ $\beta$-deformed theory for the square or $A_2^*$ lattice.
As discussed in \S.\ref{seiberg}, the amount of the global supersymmetry on a non-commutative 
lattice  and commutative one is the same. The global supersymmetry of the deformed matrix model is fewer than the undeformed theory by its construction.  Moreover, the deformed matrix model formulation has the same number of supersymmetries as  the  lattice formulation, both  can be written in terms of  identical superfields and  they possess  exactly the same supersymmetries. 

Therefore, the claim of preserving all the (global) supersymmetries in the lattice theory is identical, in the matrix model language,  to the statement that the deformation of the superpotential does not reduce the amount of supersymmetry, which is a contradiction. 

Apparently, the explicitly broken supersymmetries of the deformed matrix model are the ones 
associated with the ``link supersymmetries". The above simple argument shows that there is 
no such global supersymmetry in the theory.

{\bf An independent argument:} 
 It is also useful to reiterate what is asserted above slightly differently.  
 Again, in order not to dwell into the technical discussion on the various implementation of 
 Dirac-K\"ahler  fermions, we choose the simplest example which carry the adequate message, and phrased everything in well-known $\CN=1$ superfield language.    
Consider, for example, the $\CN=2$ SYM theory    with a gauge group $G=U(Nk)$ on $\R^4$ 
 or its dimensional reductions down  to $d < 4$.  For our conclusions, (which are  elementary), the dimension does not matter because dimensional reduction in continuum commutes with the total number of supersymmetry.  
 The $d=4$ dimensional theory possess an  $[SU(2) \times U(1)]_{\cal R}$ symmetry.    
The structure of the $\CN=2$  supersymmetry in terms of $\CN=1$ multiplets 
$V= (A_\mu, \lambda)$,  $\Phi= (\phi, \psi)$  and   $\CN=1'$ multiplets 
$V'= (A_\mu, \psi)$,  $\Phi'= (\phi, \lambda)$ 
(all in the adjoint representation of gauge group $G$)   
are shown below:
\begin{equation}
 \xymatrix{
& A_{\mu} \ar@{<->}[dl]_{\CN=1}  \ar@{<.>}[dr]^{\CN=1'}  & \\ 
\lambda \ar@{<.>}[dr]_{\CN=1'}  &  & \psi \ar@{<->}[dl]^{\CN=1}    \\
& \phi &
 }  
\end{equation}
In order to generate a one-dimensional lattice, we perform an orbifold projection 
by a $Z_k$ factor. We  assign an $r$-charge +1 to   $\Phi= (\phi, \psi)$ 
and  0 to $V= (A_\mu, \lambda)$. The result is described by a one-dimensional quiver (lattice)
with a segment 
\begin{equation} 
 \xy 
(-16,0)*{V_{n-1}}; 
(-16,0)*\xycircle(4,4){-}="V_{n-1}";
(-28,0)**\dir{-} ?(.75)*\dir{<}+(0,5)*{\scriptstyle \Phi_{n-2}}; 
(0,0)*{V_n}; 
(0,0)*\xycircle(4,4){-}="V_n"; 
(-12,0)**\dir{-} ?(.75)*\dir{<}+(0,5)*{\scriptstyle \Phi_{n-1}}; 
"V_n";(12,0)**\dir{-} ?(.75)*\dir{>}+(0,5)*{\scriptstyle \Phi_{n}}; 
(16,0)*{V_{n+1}}; 
(16,0)*\xycircle(4,4){-}="V_{n+1}"; 
"V_{n+1}";(28,0)**\dir{-} ?(.75)*\dir{>}+(0,5)*{\scriptstyle \Phi_{n+1}};
\endxy 
\end{equation}
Apparently, the supersymmetry of the quiver is only the $\CN=1$ bit, with multiplets 
$V_n= (A_{\mu,n}, \lambda_{n})$ which transform as  adjoint under the
gauge group factor $G_n$ and    $\Phi_n= (\phi_{n}, \psi_{n})\equiv(\phi_{n, n+1}, \psi_{n,n+1}) $ which transform as 
bi-fundamental under $G_n \times G_{n+1}$.  Thus, in the quiver, the  $\CN=1'$ is explicitly 
violated.  
\begin{equation}
\xymatrix{
& A_{\mu,n} \ar@{<->}[dl]_{\CN=1}   \ar@{<.>}[dr]|{\rm nothing }   & \\ 
\lambda_n    \ar@{<.>}[dr]|{\rm nothing}  &  & \psi_{n, n+1} \ar@{<->}[dl]^{\CN=1}    \\
& \phi_{n,n+1} &
 }
\end{equation}
The action of a global supersymmetry transformation of an adjoint cannot produce 
a bi-fundamental. 
According to  the interpretation of 
\cite{D'Adda:2004jb,D'Adda:2004ia, D'Adda:2005zk,D'Adda:2007ax}, 
there exist  ``link supersymmetries" which are the images of the 
$\CN=1'$ supersymmetry of  the parent.   
However, no such symmetry exists in the quiver theory or any of its dimensional reductions down to $d<4$.  

\subsection{Reinterpreting the link(2)   constructions: Why are there intriguing?}  
\label{link2r}
The  (non-supersymmetric)-lattices that are classified as  link(2) or orbifold(2) are intriguing in their own right.  They have a set of remarkable properties and below, we will describe some of them.  Some of  the interpretation we give below is in sharp contrast with 
\cite{D'Adda:2004jb,D'Adda:2004ia, D'Adda:2005zk,D'Adda:2007ax, Nagata:2007mz}. 
\begin{itemize}
{\item[\bf 1)]   Link(2) theories do not possess any global supersymmetry at the microscopic level 
 in the canonical sense. They are  $\CQ=0$ (non-supersymmetric) orbifold projections of some parent matrix theory.}

{\item[\bf 2)] Link(2) lattices  with $\CQ=0$   possess  larger  discrete point group symmetries  than 
the link(1) lattices for which  $\CQ={\rm few}$.
 The point group symmetry is in the diagonal subspace of the chiral ${\cal R}$-symmetry and Lorentz symmetry. Thus,  large discrete subgroups of the 
 chiral ${\cal R}$-symmetry are exactly realized on the lattice. }
 
 {\item[\bf 3)]Link(2) lattices provide a novel lattice structure and novel implementation of the lattice fermions which is free  of doubling, just like the staggered fermions.  
  The link(2) is not a natural implementation of the Dirac-K\"ahler decomposition. For the latter, the fermions are not all on the same footing.  }
 
  {\item[\bf 4)] The {\it classical} continuum limit of all link(2) lattices has full extended supersymmetry! }
  
 { \item[\bf 5)] In link(2) lattices, unlike link(1) or continuum, there are no gauge invariant  
 Grassmann odd observables (or 
 fermionic   operators).\footnote{Of course, this simple fact is sufficient to deduce that there is no  exact  supersymmetry in link(2) formulations. The exact supersymmetry, if it exist, maps 
 gauge invariant bosonic operators (and  states)  into  fermionic operators (and states)  or vice versa.  Since there are no Grassmann odd  observables in lattice(2) formulations, this also implies the absence of any exact supersymmetry.}}
  \end{itemize}
 
The third property implies that the classical spectrum of propagating 
fermionic  and bosonic  fields coincide despite the absence of any exact supersymmetry on the lattice. For all link(1) or link(2) type cubic 
lattices, we obtain 
\beq 
{\cal P}_{\mu} \equiv \frac{2}{a} \sin \frac{a p_{\mu}}{2}, \qquad (M_{\bf p}^{\rm fermions})^2 =  (M_{\bf p}^{\rm bosons})^2 = \sum_{\mu=1}^{d} {\cal P}_{\mu}^2
\eqn{spec}
\eeq
for (on-shell) degrees of freedom.   In \Eq{spec}, $p_{\mu}= \frac{2 \pi}{Na} n_{\mu}, n_\mu =0, 1, \ldots, N-1$ is the momenta in the Brillouin zone. 
Thus, at the classical level, these theories (regardless of whether one starts with link(1) or link(2) 
formulations), produce  a Lorentz invariant continuum theory with full extended 
supersymmetry of the 
continuum!  This does not mean that microscopic theory has the full supersymmetry. \footnote{This remarkable property leads to some  misinterpretation in literature. It is sometime stated that exact supersymmetry is realized classically (with modified Leibniz rule etc \ldots) , and one needs to check it at quantum level, after radiative corrections are taken into account. Assuming the first statement is correct, the latter would be an analysis of 
 the spontaneous breaking/nonbreaking  of supersymmetry.
 What happens in reality in link(2) theories is following:  At the cut-off, there is no supersymmetry. At tree level (classical) continuum,   there is an emergent full set of supersymmetry.   However, whether this tree level conclusion is true or not quantum mechanically depends on the radiative corrections. In order to answer the latter (at least in perturbation theory), one needs to check all the relevant and marginal operators allowed by microscopic symmetries, and then check, whether they are generated or not. 
If there are no such dangerous operators, then the classical result is also valid in quantum continuum limit, and one has continuum supersymmetry without any microscopic supersymmetry.   If there are dangerous relevant operators which do get generated,  then quantum continuum limit is non-supersymmetric as the microscopic theory. Then, 
one needs to fine-tune  to recover supersymmetry in the continuum. 
}

The  degeneracy of fermions and bosons at the classical level (despite the absence  of exact supersymmetry)   should not be viewed as a surprise. The spectral degeneracy and the absence of  doublers is  an aspect of the {\it structure }
of these lattices, not supersymmetry, just like staggered fermion or geometric 
Dirac-K\"ahler fermions.



 A more  important questions is whether the symmetries of these lattices and  spectral degeneracy  of fermions and bosons    can be used to reduce the fine tunings to achieve the desired quantum continuum limit. This is an issue in which naive arguments may fail. This will be discussed in a specific example [$\CN=(2,2)$ theory] at the end of next subsection.

\subsection{Representation theory of link(2) lattices and Dirac-K\"ahler fermions}  
As emphasized in item 2) and 3), the link(2) 
formulations present novel implementations of lattice fermions, reminiscent of staggered fermions and Dirac-K\"ahler fermion.  In \S\ref{sec:twisting}, we will review the precise relation between staggered fermions and  Dirac-K\"ahler fermions (or  twisting) from the viewpoint of symmetries, as  in Fig.\ref{godd}.  Below, we  discuss at the level of representation theory, the relation between 
link(2) formulation and  Dirac-K\"ahler fermions.   Two examples will be detailed, the
$d=2$ dimensional link(2) formulation of $\CN=(2,2)$ target theory  \cite{D'Adda:2005zk} 
with dihedral $D_4$ point group symmetry of order  $|D_4|=8$  and the $d=3$ dimensional link(2) formulation of $\CN=4$ target theory  \cite{D'Adda:2007ax} with  full octahedral symmetry $O_h$ 
with order $|O_h|=48$.  The  generalization to the other link(2) theories is obvious.    Recall that for the supersymmetric lattices  with $\CQ=1$  
\cite{Cohen:2003xe, Cohen:2003qw} has much smaller, order $|Z_2|=2$ and $|S_3|=6$ respectively. The main point that we wish to emphasize is that the exact supersymmetry in the 
formulations of \cite{Cohen:2003xe, Cohen:2003qw} is traded with much larger point 
group symmetry of  the $\CQ=0$ lattices of  \cite{D'Adda:2005zk,D'Adda:2007ax}.

As  emphasized with  Fig.\ref{godd},  the point group symmetries  for orbifold  and  link approach  
lattices 
should not be interpreted as being subgroups of ordinary Lorentz group, rather they are  the subgroups living in  the diagonal sum of the chiral ${\cal R}$-symmetry and Lorentz group, i.e, $G_{\rm point}' \subset SO(d)'$.  Below, we analyze the representation theory of $G_{\rm point}$ for the two examples.  
 Since exact chiral symmetry is rather important in preventing certain dangerous relevant 
and marginal operators, and link(2) formulations has very large discrete chiral symmetries, 
the following analysis is useful  in studying the quantum continuum limit of these theories.

\subsubsection{A  $\CQ=0$ link(2) lattice and    twisted dihedral group}

\begin{figure}
\begin{center}
\includegraphics[angle=-90,width=4 in]{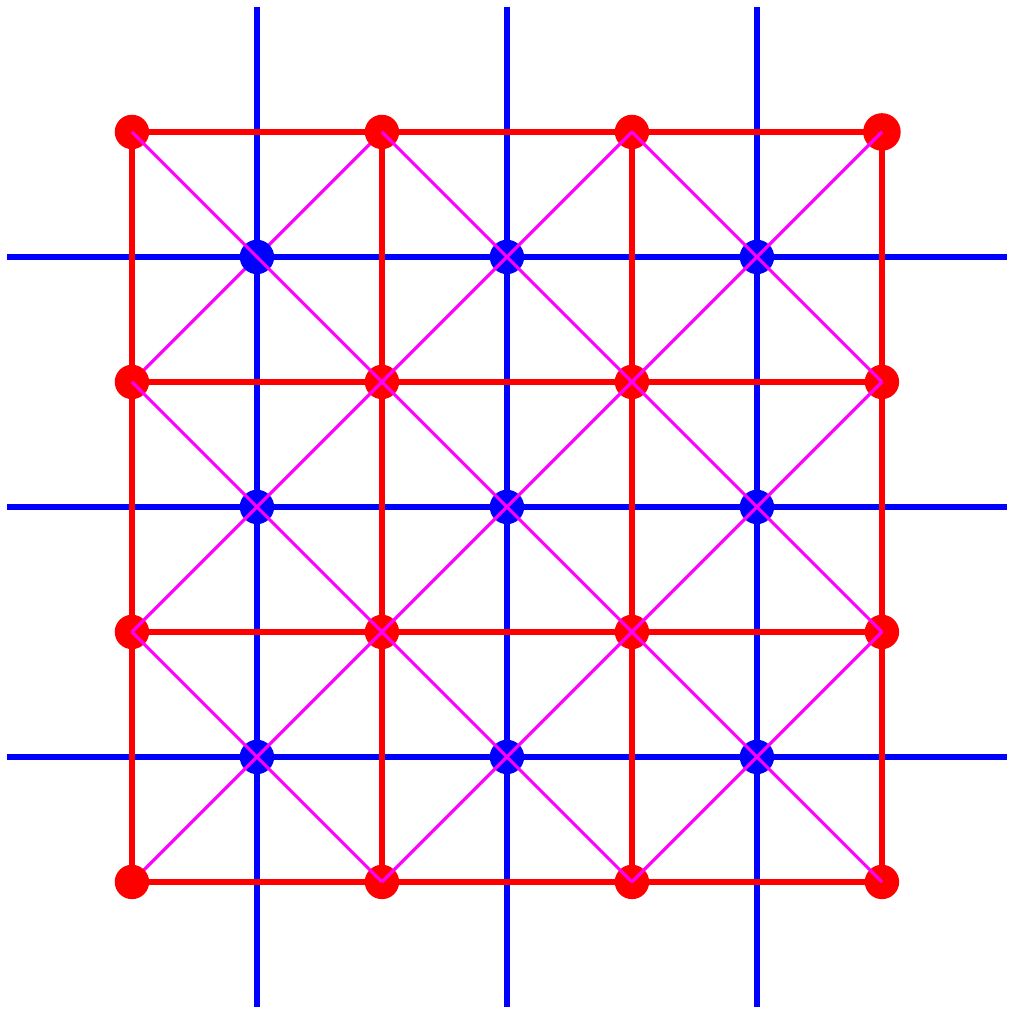}
\caption
    {\small
The two dimensional  lattice structure of the $\CQ=0$ link(2) lattice formulation of $\CN=(2,2)$  
target theory.  The lattice can be split into even (red) and odd  (blue) sublattices.  The bosonic degrees of freedom reside on red and blue links.  The fermions reside on the diagonal magenta links.  This lattice  theory can be obtained by a non-supersymmetric  orbifold projection of the 
$\CN=(2,2)$ matrix model. There are no dynamical fields residing on the sites.
 }
\label{linklat}
\end{center}
\end{figure}

 The matter content of the $\CQ=0$ link(2) lattice for the $\CN=(2,2) $ SYM  target theory  
is as follows: On a unit cell, there are two types of  complexified bosonic   fluctuations $(z_1, z_2)$  (and their conjugates) and 
four  types of Grassmann fields 
$  \alpha_{12}, \alpha_{\bar 12},  \alpha_{1\bar 2},  \alpha_{\bar 1 \bar 2}   $. The fermions and bosons are associated with links:
 \beq 
&& \alpha_{12} (\bfn): {\bfn} \rightarrow \bfn +\bfe_1 + \bfe_2, \qquad  \alpha_{\bar 12}  (\bfn): \bfn \rightarrow 
 \bfn -\bfe_1 + \bfe_2, \qquad \cr
&&  \alpha_{1\bar 2}( \bfn): { \bfn} \rightarrow  \bfn +\bfe_1 - \bfe_2, \qquad  
\alpha_{\bar 1 \bar 2}( \bfn): { \bfn} \rightarrow  \bfn - \bfe_1  - \bfe_2, \qquad   \cr
&&  z_{1}( \bfn ):  { \bfn} \rightarrow   \bfn +2\bfe_1  \qquad  \qquad  z_{2}( \bfn) :   \bfn 
\rightarrow  \bfn +2\bfe_2  \cr 
 &&  \bar z_{1}( \bfn ): \bfn +2\bfe_1 \rightarrow  { \bfn}    \qquad  \qquad  
 \bar z_{2}( \bfn) :   \bfn  +2\bfe_2  
\rightarrow \bfn
 \eeq
 where $\bfn$ is site index, $(\bfe_m)_{n}= \delta_{m n} $ for   $m,n=1, 2$.
 The highly symmetric structure of the lattice is shown in Fig.\ref{linklat}. This lattice can be obtained by an orbifold projection which preserves none of the supersymmetries 
 \cite{Kaplan:2002wv}. The ${\bf r}$-charge assignments are ${\bf r}(z_1)=(2,0), {\bf r}(z_2)=(0,2),  {\bf r}(\alpha_{12})=(+1, +1), {\bf r}(\alpha_{\bar 12})=(-1, +1)$ etc, 
 and this is in 
  a  one-to-one mapping with  the position of the lattice  fields on a unit cell.     
 The action is 
\beq
S^{\rm link(2)} = &&\frac{1}{g^2} \sum_{\bfn} \Tr  \Big[ \half \Big(   \bar z_1(\bfn-2\bfe_1)  z_1( \bfn - 2\bfe_1)  -  z_1( \bfn) 
\bar z_1( \bfn)   + (1 \leftrightarrow 2) \Big)^2   \cr  \cr 
&&
 + 2 \Big| z_1(\bfn)   z_2(\bfn+ 2\bfe_1)  -  z_2(\bfn)  z_1(\bfn+ 2\bfe_2)  \Big|^2  
 \cr  \cr
&& +  \sqrt 2 \left(  \Delta_\bfn (\alpha_{12}, \bar z_1, \alpha_{1 \bar 2})  + 
 \Delta_\bfn ( \alpha_{12} , \bar z_2,  \alpha_{\bar 1 2} )    +  \Delta_\bfn ( \alpha_{\bar 1 \bar 2} , 
  z_1,  \alpha_{\bar 1  2} )   -  \Delta_\bfn ( \alpha_{\bar 1\bar 2},   z_2,  \alpha_{1 \bar 2} ) \right)
 \Big]  \qquad \qquad 
 \eqn{link2a}
\eeq 
where we have used the triangular plaquette function $\Delta_\bfn$ given by 
\beq
\Delta_\bfn (\alpha_{12}, \bar z_1, \alpha_{1 \bar 2}) =  
  \alpha_{12}(\bfn)  \Big(\bar z_1 (\bfn-  \bfe_1 + \bfe_2)  \alpha_{1\bar 2}  (\bfn-  \bfe_1 + \bfe_2)   - 
   \alpha_{1\bar 2}  (\bfn+   \bfe_1 + \bfe_2 )
\bar z_1(\bfn)    \Big)  \qquad \qquad  \cr \cr
\Delta_\bfn (\alpha_{12}, \bar z_2, \alpha_{\bar 1 2}) =  
  \alpha_{12}(\bfn)  \Big(\bar z_2 (\bfn+  \bfe_1 - \bfe_2)  \alpha_{\bar 1 2}  (\bfn +  \bfe_1 - \bfe_2)   - 
   \alpha_{\bar 1  2}  (\bfn+   \bfe_1 + \bfe_2 )
\bar z_2(\bfn)    \Big) \qquad \qquad   \cr \cr
\Delta_\bfn (\alpha_{\bar 1 \bar 2},  z_1, \alpha_{\bar 1 2}) =  
  \alpha_{\bar 1\bar 2}(\bfn)  \Big( z_1 (\bfn-  \bfe_1 - \bfe_2)  \alpha_{\bar 1 2}  (\bfn +  \bfe_1 - \bfe_2)   - 
   \alpha_{\bar 1  2}  (\bfn -  \bfe_1 - \bfe_2 )
 z_1(\bfn - 2\bfe_1 )    \Big)  \qquad  \cr  \cr
 \Delta_\bfn (\alpha_{\bar 1 \bar 2},  z_2, \alpha_{ 1 \bar 2}) =  
  \alpha_{\bar 1\bar 2}(\bfn)  \Big( z_1 (\bfn-  \bfe_1 - \bfe_2)  \alpha_{1 \bar 2}  (\bfn -  \bfe_1 + \bfe_2)   - 
   \alpha_{ 1 \bar  2}  (\bfn -  \bfe_1 - \bfe_2 )
 z_2(\bfn - 2\bfe_2 )    \Big)  \qquad 
\eeq
This is the link(2) action studied in   \cite{D'Adda:2005zk}.
 In the  discussion of the representation theory of point group symmetry,  we ignore lattice site index  $\bfn$ for convenience. \footnote{\label{QCD} 
 {\bf A novel lattice formulation for QCD(adj):} Slight modification of this action can also be used in formulating QCD with adjoint fermions in  various dimensions.     Substitute  complex bosonic link matrices $z_m(\bfn)$ with  group valued unitary link matrices $U_m(\bfn)$.   Resulting theory is a new lattice formulation  of lattice QCD(adj) in two dimensions.
 Generalization to $d=4$ dimensions is obvious,     and is an alternative for  staggered fermions. To obtain  QCD(adj) with four Weyl fermions from   the link(2) $\CN=4$ SYM, use the prescription:   $z_m(\bfn) \rightarrow \delta_{m\mu} U_{\mu}(\bfn) +  \delta_{m5 } 0,$ where  $ 
   m=1, \ldots, 5, \;  \mu=1, \ldots 4   $ which replaces 
four   algebra valued fields with the group valued once and set the extra scalar to zero.  
 }

The $\CQ=0$  link(2) formulation of the $\CN=(2,2)$ has a $D_4$ point group symmetry with  order eight.   These are the full set of symmetry operations of a square and are shown in 
Table.\ref{tab:D4}.   
\setlength{\extrarowheight}{-3pt}
\begin{table}[t]
\centerline{
\begin{tabular}
{|c|c|c|c|c|c|} \hline
classes: & ($e$)  & (2$C_4$) & ($C_2$) & (2 $C_2'$)&(2$C_2''$) \\ \hline
$A_1$ & 1  &   1  &   1   &  1    &   1        \\ 
$A_2$ & 1  &  1  &   1   & -1    &   -1       \\ 
$B_1$ & 1 &   -1  &   1   &  1   &  -1           \\
$B_2$ & 1 &  -1  &   1   &  -1    &  1        \\ 
$E$ & 2  &   0  &   -2   &  0    &   0    \\ \hline
\end{tabular} }
\caption{\sl The character table of $D_4$, the point symmetry group of
two dimensional  link(2)  lattice formulation for $\CN=(2,2)$ theory   \cite{D'Adda:2005zk}.
$e$ is identity, $C_4$ and  $C_2$ are rotations by $\pi/4$ and $\pi/2$, 
 $C_2'$ are reflections with respect to $\bfe_1$ and $\bfe_2$ axis, and 
  $C_2''$ are reflections with respect to diagonals, $\bfe_1 \pm \bfe_2$.
  \label{tab:D4}}
\end{table}
This should be contrasted  with the $Z_2$ point group symmetry of   the $\CQ=1$ supersymmetry preserving regularization of the $\CN=(2,2)$  SYM theory  
\cite{Cohen:2003xe}. The apparent trade-off here is between supersymmetry and point group symmetry.  In link(2), one achieves much larger point group symmetry $D_4$ to the price of 
giving up the exact lattice supersymmetry.

 One other interpretational distinction relative to  \cite{D'Adda:2005zk}  that  
  we wish to emphasize is that, the link fermions are not 
the natural implementation of the Dirac-K\"ahler decomposition on lattice (although see the appendix of  \cite{D'Adda:2004jb}). In particular, 
in link(2) formulation, all the fermions are on the same footing.
They do transform to one another  under $\pi/4$ rotations, and this is also an 
invariance of action.   However, in a 
natural implementation of Dirac-K\"ahler fermions, the fermions are one zero form $\psi^{(0)}$, two one-form  $\psi^{(1)}$ and one two-form   $\psi^{(2)}$ . Obviously, no lattice rotation can map a zero form $\psi^{(0)}$ to a one-form $\psi^{(1)}$ or vice-versa.
Therefore, in what sense, the link(2) formulation  is related to the Dirac-K\"ahler fermions? 
One other related puzzle: Obviously, $\psi^{(0)}$  is a scalar under $SO(2)'$. Therefore, 
it {\bf must be} in a scalar representation of any  discrete subgroup  $G_{\rm point}$ of 
$SO(2)'$.  How does this reconcile with the link nature of all the fermions?  
 
In order to answer these questions, we  classify the fields on the lattice in terms of the irreducible representations of   $D_4$.
We expect the irreducible representations under  $D_4$ to have a natural interpretation under 
$SO(2)'$.  To do so, we consider 
the action of the elements of $D_4$  (one from each conjugacy class)
on the lattice fields, and then evaluate the character of the operation.  The action of 
$g \in  D_4$ on elements of a unit cell is given by  
\begin{eqnarray}
&&(e) :  (  \alpha_{12}, \alpha_{\bar 12},  \alpha_{\bar 1\bar 2},  \alpha_{1 \bar 2} ),  \;\;  (z_1, z_2  ) \rightarrow  (  \alpha_{12}, \alpha_{\bar 12},  \alpha_{\bar 1\bar 2},  \alpha_{ 1 \bar 2} ),  \;\;  ( z_1, z_2  ) 
 \cr  \cr
&& C_4 : \qquad \qquad \qquad \qquad \qquad \qquad \;\;  \rightarrow   (   \alpha_{\bar 12},  \alpha_{\bar 1\bar 2},  \alpha_{ 1 \bar 2},  \alpha_{12} ),  \;\;  (z_2, \bar z_1  ) 
\cr \cr
&& C_2 : \qquad \qquad \qquad \qquad \qquad \qquad  \;\; \rightarrow   (   \alpha_{\bar 1\bar 2},  \alpha_{ 1 \bar 2},  \alpha_{12} ,  \alpha_{\bar 12}
),  \;\;  (\bar z_1, \bar z_2              )  \cr \cr
&& C_2' : \qquad \qquad \qquad \qquad \qquad  \qquad \;\; \rightarrow   (     
 \alpha_{ 1 \bar 2},  \alpha_{\bar 1\bar 2}, \alpha_{\bar 12}, \alpha_{12}
),  \;\;  (z_1, \bar z_2          )  \cr \cr 
&& C_2'' : \qquad \qquad \qquad \qquad \qquad \qquad \;\; \rightarrow   (    
 \alpha_{12},   \alpha_{1 \bar 2},   \alpha_{\bar 1\bar 2},\alpha_{\bar 12}
),  \;\;  ( z_2, z_1            )  
\end{eqnarray}
The  character is $\chi(g)= \Tr( M(g))$, where  $M(g)$ is a 
matrix representation of the operation $g$.   
Since the  character is a class function, it is independent of 
representative.  Thus, we make a character multiplet   $\left[ \chi(M(e)), \ldots, \chi( M(C_2'') )\right]$. 
For the fermions, the hermitian $\Re({\rm bosons})$ and  anti-hermitian $\Im({\rm bosons})$ components of link bosons,   we obtain  
\beq
&&\chi^{\rm fermions} =   [4,0,0,0,2] = A_1 \oplus  B_2 \oplus E \cr \cr 
&&\chi^{\Re(\rm bosons)} =   [2,0,2,2,0] = A_1 \oplus  B_1 \cr \cr
&&\chi^{\Im(\rm bosons)} =   [2,0,-2,0,0] = E 
\eqn{D4splitting}
\eeq
 The gauge bosons [$\Im({\rm bosons})$] and scalars [$\Re({\rm bosons})$] respectively fill in vector and pseudo-vector  representation of the twisted $SO(2)'$.    Under the $D_4$ subgroup, 
 there is a two dimensional irreducible  representation  
 $E$  corresponding to vectors. The pseudo-vector is reducible and splits as 
    $A_1 \oplus  B_1$.
The fermions apparently form a reducible representation and split into two one dimensional representations ($A_1$ and $B_2$) and a two dimensional vector representation $E$. 
We can indeed identify the irreducible representations of $D_4$ with the natural realization of the 
Dirac-K\"ahler fermions on the lattice, for example, the zero-form 
$\psi^{(0)} \sim  \frac{1}{2} ( \alpha_{12}+  \alpha_{\bar 12} + \alpha_{\bar 1\bar 2} + \alpha_{1 \bar 2} ) $ where the right hand side corresponds to $A_1$.   Thus, the irreducible representation of the 
$D_4$ nicely maps into the  Dirac-K\"ahler twisted version of continuum, with twisted 
rotation group $SO(2)'$. In other words, 
\beq
A_1 \oplus E \oplus  B_2  \sim  \psi^{(0)} \oplus  \psi^{(1)} \oplus  \psi^{(2)} \sim 
   1 \oplus 2 \oplus 1 
\eeq
Similar phenomena also takes place in supersymmetric $A_d^{*}$  lattices  
 where the decomposition of link and face fermions 
into the  irreducible representations under the permutation group $S_{d+1}' \subset SO(d)'$ results in the usual  Dirac-K\"ahler decomposition. 
This is also how   the link(2) lattice produces the 
Dirac-K\"ahler twist in its continuum.  

{\bf Remark:}  The gauging of the Dirac-K\"ahler fermions and link fermions are also different.  For example, although 
the $\frac{1}{2} ( \alpha_{12}+  \alpha_{\bar 12} + \alpha_{\bar 1\bar 2} + \alpha_{1 \bar 2} ) $ is a singlet under $D_4$, it {\bf cannot} be contracted with any other  $D_4$ singlet 
to form a relevant (or irrelevant) gauge singlet operator at any finite lattice spacing.  
The reason is, in the gauged lattice theory, $( \alpha_{12}+  \alpha_{\bar 12} + \alpha_{\bar 1\bar 2} + \alpha_{1 \bar 2} )$ does  not transform co-variantly  under gauge rotations.  Let $G(\bfn) $ denote a gauge rotation associated with  site $\bfn$. Then, under a gauge transformation, the 
constituents transform as  
\beq
 \alpha_{12}(\bfn) \rightarrow G(\bfn) \; \alpha_{12}(\bfn) \; G^{\dagger} ({\bfn + \bfe_1 + \bfe_2}), \qquad 
  \alpha_{\bar 12}(\bfn)  \rightarrow G(\bfn) \; \alpha_{\bar 12}(\bfn) \; G^{\dagger}( \bfn - \bfe_1 + \bfe_2),\cr 
 \alpha_{1\bar 2}(\bfn) \rightarrow G(\bfn) \; \alpha_{1\bar 2}(\bfn) \; G^{\dagger}(\bfn + \bfe_1 - \bfe_2) ,\qquad 
  \alpha_{\bar 1 \bar 2}(\bfn) \rightarrow G(\bfn) \; \alpha_{\bar 1 \bar 2} (\bfn) \; G^{\dagger}({\bfn - \bfe_1 - \bfe_2}).
\eeq
This means, the combination of gauge invariance and $D_4$ symmetry is vastly more restrictive than  each would be individually.  
The combination  restricts the type of operators that one can write down.  This also shows that, at finite lattice spacing, the link fermions and Dirac-K\"ahler implementation are truly different. 
For example, the zero form site fermion transform by conjugation, $\psi^{(0)}  (\bfn) 
\rightarrow      G(\bfn) \; \psi^{(0)} (\bfn) \; G^{\dagger} ({\bfn })$. 
 Of course, 
 in classical continuum limit, this difference is lifted,  since all the fermion and boson fields  transform as adjoints.

{\bf Comments  on classical and quantum continuum limits:}    
Consider the classical and quantum continuum limit of the link(2) formulation. In the classical continuum limit, 
\beq 
S^{\rm link(2)} = S^{\CN=(2,2)} [1 + O(aq)]
\eqn{classical}
\eeq
where  $S^{\CN=(2,2)}$ is the action for the continuum $\CN=(2,2)$ theory, and $q$ is some Euclidean momenta.  

We wish to understand the quantum continuum limit of these theories when the radiative 
corrections are taken into account.  (Below, we follow the analysis of \S5 of Ref.\cite{Cohen:2003xe} verbatim.)
Consider a radiative correction to the action of an operator ${\cal O}$ with dimension $p$
\beq
\delta S= \frac{1}{g_2^2} \int d^2x C_{\cal O}  {\cal O}
\eeq 
Since the lattice theory is a $\CQ=0$ theory, there is no integration over a  superspace coordinate.   In  power counting, we use the classical scaling dimensions, $[dx]=-1, \; 
 [{\rm bosons}]=+1, \;  [{\rm fermions}]=+3/2, \; [g_2^2]= +2$, and $a$ is the lattice spacing.  
 By \cite{Cohen:2003xe}, the coefficient  $C_{\cal O}$  has a loop expansion 
\beq
 C_{\cal O}  = a^{p-4} \sum_{\ell} c_{\ell} (g_2^2 a^2)^{\ell}
\eqn{opan}
\eeq 
where  $c_{\ell}$ may have logarithmic dependence on the lattice spacing $a$. 

The operators for which $p-4 + 2 \ell \leq 0$ are the only possible local counter-terms. At classical $\ell=0$ level, the  long distance action for the lattice theory agrees with the target theory, as shown in \Eq{classical}. For $l\geq 2$,    there are no local relevant or marginal counter-terms  that get induced radiatively.  However, for $\ell =1$, the scalar mass operator 
with $p=2$ may receive a logarithmic correction. 

This is unlike the  $\CQ=1$ supersymmetric lattice   \cite{Cohen:2003xe}. In that case,  only the counter-terms with    $p-7/2 + 2 \ell \leq 0$ are  possible due to exact supersymmetry and the scalar mass operator  does not get induced radiatively.

The scalar mass operator  is a relevant operator which does affect the physics of the target theory, and it is allowed by all the symmetries of the link(2) lattice action.  Is it, however,  possible that an operator which is allowed by all the symmetries of the microscopic theory may not be generated? Or is there a reason to think that the behavior of these theories in the continuum may be tamer than the above analysis suggests? 
Naively,   the spectrum of fermions and bosons are {\it degenerate} even at a finite lattice spacing, and the number of degrees of freedom of both types is {\it balanced}. 
For each fermionic loop, there is a bosonic loop and vice versa. 
 Moreover, the \Eq{classical} 
implies that  the interaction vertices of the theory defined by $S^{\rm link(2)}$, close to the continuum limit,  may be expressed as 
\beq 
V^{\rm lattice}= V^{\rm cont.}[1 +O(qa)]
\eeq 
When inserted into loops, the leading term really just behaves like the extended supersymmetric 
theory and the correction has an extra suppression factor relative to  \Eq{opan}. 
Perhaps, despite the absence of 
any supersymmetry at the cut-off, these features may be sufficient to suppress dangerous relevant  operators.  That would be another way to have naturally light scalars without microscopic supersymmetry or shift symmetry, and would be remarkable.  

However, the above line of reasoning may be too naive.   In a lattice gauge theory and effective theories, there are cases in which a naively irrelevant operator becomes important and generates  unwanted relevant operators.  Such behavior may occur
  if the lower dimension relevant operator is not protected by a symmetry.  
  The best known example, which has a resemblance to the   above discussion, is about 
  the chiral symmetry on the lattice and Wilson fermions \cite{Kaplan:2007zz}.
 \footnote{I thank David B. Kaplan for the line of reasoning below.} 
 The Wilson's  lattice fermion Lagrangian is 
 \beq
 \mybar \psi (i \gamma_{\mu} D_{\mu} -m -a r \Delta) \psi
 \eeq
where $D_{\mu}$ and $\Delta$ are gauge covariant Dirac-operator and Laplacian, $a$ is lattice spacing, $m$ is bare mass and $r$ is an order one parameter introduced to lift the spurious doublers. In the naive continuum limit, the operator proportional to lattice spacing is an irrelevant dimension five operator.  Both $ m$ and $r$ terms explicitly violate the chiral symmetry.  In this theory, the fermion mass term, instead of being multiplicatively renormalized, is additively renormalized by a term proportional to $r/a$.  Thus, the naively irrelevant dimension five operator 
radiatively induces a dimension three operator.   If the target theory is massless or a theory with a light fermion, the exact or approximate chiral symmetry of the naive continuum limit is spoiled by a so called  ``irrelevant" operator.

The danger in the link(2) formulation is analogous.  There is no symmetry which protects scalar masses in these formulations in general. The naive classical continuum limit has supersymmetry. What one really needs to check are the higher dimension, irrelevant 
operators which may generate the scalar mass operator when inserted into loops.  Perhaps,  
just like the absence of the chiral symmetry  does not admit  naturally light fermions, the absence of the exact supersymmetry does not admit light scalars either. \footnote{However, 
both supersymmetric link(1) and non-supersymmetric link(2) formulations are the orbifold projections of a supersymmetric 
 matrix model.  
There is a non-perturbative equivalence between parent-daughter pairs related to one another by orbifold projections. The necessary and sufficient conditions for the validity of these large $N$ equivalences can be found in  \cite{Kovtun:2003hr}. In particular, such large $N$ equivalences 
imply the daughter-daughter equivalences, in some cases relating a supersymmetric theory
to  a non-supersymmetric one.   In particular, link(1) and link(2) formulations are  such pairs. 
In link(1), scalar mass term is forbidden by supersymmetry. 
The equivalence implies, if the  mass term is generated for scalars in link(2), it must be an  
$O(1/N)$ effect. In phenomenology, 
in a class of non-supersymmetric  theories,  Ref.  \cite{Strassler:2003ht}
 argued the existence of light  scalars and large hierarchies without fine-tuning as a consequence of such  susy-nonsusy daughter-daughter equivalences.  It is likely that similar suppression of various dangerous operators may also take place in link(2) theories, at least in the large $N$ limit. These observations are in agreement  
 with the structure of the perturbative planar and non-planar loop expansions discussed by 
 Nagata \cite{Nagata:2008zz}.}
To sum up,  we are inconclusive about the  amount of fine-tuning in the 
quantum continuum limit of the $\CQ=0$ link(2) theory.

 \subsubsection{Representation theory for the twisted full octahedral group}
 The matter content of the $\CQ=0$ link(2) lattice for the $d=3$ dimensional $\CN=4 $ SYM  target theory   is as follows: On a unit cell, there are three  types of  complexified bosonic   fluctuations $(z_1, z_2,z_3)$  (and their conjugates) and 
eight   types of Grassmann fields 
$  \alpha_{123}, \alpha_{\bar 123}, \ldots  $  etc. 
The fermions  reside on the links 
 \beq 
&& \alpha_{123}: {\bf 0} \rightarrow  +\bfe_1 + \bfe_2 + \bfe_3 \qquad  \alpha_{\bar 1 \bar 2 \bar 3}: {\bf 0} \rightarrow 
 -\bfe_1 - \bfe_2 - \bfe_3 \qquad \cr
  && \alpha_{1\bar 2\bar 3}: {\bf 0} \rightarrow  +\bfe_1  - \bfe_2 -  \bfe_3 \qquad  \alpha_{\bar 1  2  3}: {\bf 0} \rightarrow 
 - \bfe_1 + \bfe_2 + \bfe_3 \qquad \cr
&& \alpha_{\bar12\bar3}: {\bf 0} \rightarrow  -\bfe_1 + \bfe_2  - \bfe_3 \qquad  
\alpha_{ 1 \bar 2  3}: {\bf 0} \rightarrow 
 +\bfe_1 - \bfe_2 + \bfe_3 \qquad \cr
&& \alpha_{\bar  1 \bar 23}: {\bf 0} \rightarrow  -\bfe_1  - \bfe_2 + \bfe_3 \qquad  \alpha_{ 1  2 \bar 3}: {\bf 0} \rightarrow 
 +\bfe_1 + \bfe_2 - \bfe_3 \qquad 
 \eeq
 and the bosons are associated with 
\beq
  z_{1}:  {\bf 0} \rightarrow  +2\bfe_1  \qquad  \qquad  z_{2}:  {\bf 0} \rightarrow  +2\bfe_2  \qquad  \qquad  z_{3}:  {\bf 0} \rightarrow  +2\bfe_3 , 
\eeq

The point group symmetry is the full octahedral group $O_h = O \ltimes I$ where $O$ is the pure 
rotations and $I$ is the inversion. Hence,  $O_h $ has both proper and improper rotations. The 
48 group operations and the character table are shown  in Table.\ref{tab:Oh}. 
  \setlength{\extrarowheight}{-3pt}
\begin{table}[t]
\centerline{
\begin{tabular}
{|c|c|c|c|c|c||c|c|c|c|c|c|} \hline
classes: & ($e$)  & (8$C_3$) & ($3C_2$) & (6 $C_2'$)&(6$C_4$) 
& ($i$)  & (8$S_3$) & ($3S_2$) & (6 $S_2'$)&(6$S_4$)   \\ \hline
$A_{1g}$ & 1  &   1  &   1   &  1    &   1  & 1  &   1  &   1   &  1    &   1       \\ 
$A_{2g}$ & 1  &  1  &   1   & -1    &   -1   & 1  &  1  &   1   & -1    &   -1      \\ 
$E_{g}$ & 2 &   -1  &   2   &  0   &  0  &   2 &   -1  &   2   &  0   &  0           \\
$T_{1g}$ & 3 &  0  &   -1   &  -1    &  1      & 3 &  0  &   -1   &  -1    &  1     \\ 
$T_{2g}$ & 3  &   0  &   -1   &  1   &   -1 & 3  &   0  &   -1   &  1   &   -1   \\ \hline
$A_{1u}$ & 1  &   1  &   1   &  1    &   1  & -1  &  - 1  &  - 1   &  - 1    &  - 1       \\ 
$A_{2u}$ & 1  &  1  &   1   & -1    &   -1   & -1  &  -1  &   -1   & 1    &   1      \\ 
$E_{u}$ & 2 &   -1  &   2   &  0   &  0  &   -2 &   1  &   -2   &  0   &  0           \\
$T_{1u}$ & 3 &  0  &   -1   &  -1    &  1      & -3 &  0  &   1   &  1    &  -1     \\ 
$T_{2u}$ & 3  &   0  &   -1   &  1   &   -1 & -3  &   0  &   1   &  -1   &   1   \\ \hline
\end{tabular} }
\caption{\sl The character table of full octahedral group $O_h$, the point symmetry group of the 
three  dimensional  link(2)  lattice formulation for $\CN=4$ (or $\CQ=8$) target theory   \cite{D'Adda:2007ax}. $e$ is identity, $8C_3$  are rotations by $2\pi/3$  along the body-diagonals, $3C_2$ and  $6C_4$   are rotations by $\pi/2$ and $\pi/4$ along the line passing through the center of faces, $6C_2'$ are rotations by $\pi$ along the lines cutting the edges in the middle. $i$ is inversion, and $S = C \times i$.   \label{tab:Oh} The character table of $O_h$ can be deduced from the product of octahedral group $O$ (upper-left five by five block) and the $I$ inversion group. }
\end{table}

 As in the two dimensional example, we wish to understand the representations of various lattice fields and decompose them into   their irreducible representations. It is sufficient to first inspect the   action of $g \in O$  subgroup of $O_h$ on the fields on a unit cell 
 \begin{eqnarray}
&&(e)  \;\;\;\;  :\rightarrow  (  \alpha_{123}, \alpha_{ 1\bar 2 \bar 3 },  \alpha_{\bar 1 2 \bar 3},  \alpha_{ \bar 1 \bar 2 3},     \alpha_{\bar 1\bar 2\bar 3}, \alpha_{ \bar 1 2  3 },  \alpha_{1 \bar  2 3 },  \alpha_{  1  2 \bar 3} ),  
  \;\;  ( z_1, z_2, z_3  ) 
 \cr  \cr 
&&(8C_3) :\rightarrow  (  \alpha_{231}, \alpha_{ 2\bar 3 \bar 1 },  \alpha_{\bar 2 3 \bar 1},  \alpha_{ \bar 2 \bar 3 1},     \alpha_{\bar 2\bar 3\bar 1}, \alpha_{ \bar 2 3  1 },  \alpha_{2 \bar  3 1 },  \alpha_{  2 3 \bar 1} ),  
  \;\;  ( z_2, z_3, z_1  ) 
 \cr \cr
&&(3C_2) :\rightarrow  (  \alpha_{\bar 1\bar 2 3}, \alpha_{ \bar 1  2 \bar 3 },  \alpha_{ 1 \bar 2 \bar 3},  \alpha_{ 1  2 3},     \alpha_{ 1  2\bar 3}, \alpha_{  1 \bar 2  3 },  \alpha_{\bar 1   2 3 },  \alpha_{  \bar 1  \bar  2 \bar 3} ),  
  \;\;  ( \bar z_1, \bar z_2, z_3  ) 
 \cr \cr
&&(6C_2') :\rightarrow  (  \alpha_{21\bar 3}, \alpha_{ 2\bar 1  3 },  \alpha_{\bar 2 1  3},  \alpha_{ \bar 2 \bar 1 \bar 3},     \alpha_{\bar 2\bar 1  3}, \alpha_{ \bar 2 1 \bar  3 },  \alpha_{2\bar  1 3 },  \alpha_{  2 1  3} ),  
  \;\;  ( z_2,  z_1, \bar  z_3  ) 
 \cr \cr
&&(6C_4) :\rightarrow  (  \alpha_{2\bar 13}, \alpha_{ 2  1  \bar 3 },  \alpha_{\bar 2 \bar 1 \bar 3},  \alpha_{ \bar 2 1  3},     \alpha_{\bar 2  1 \bar 3}, \alpha_{ \bar 2 \bar 1  3 },  \alpha_{2   1 3 },  \alpha_{  2  \bar 1  \bar 3} ),  
  \;\;  ( z_2, \bar z_1,  z_3  ) 
  \eqn{O}
\end{eqnarray}
The inversion  $(i)$ acts as 
 \begin{eqnarray}
&&(i)  \;\;\;\;  :\rightarrow  (  \alpha_{\bar 1\bar 2\bar 3}, \alpha_{ \bar 1  2  3 },  \alpha_{ 1 \bar 2  3},  \alpha_{  1 2 \bar 3},     \alpha_{ 1 2 3}, \alpha_{ 1 \bar  2 \bar   3 },  \alpha_{\bar 1   2  \bar 3 },  \alpha_{ \bar  1 \bar  2 3} ),  
  \;\;  ( \bar z_1, \bar z_2, \bar z_3  ) 
  \eqn{I}
 \eeq
Since $O_h = O \ltimes I$,   the character multiplet $[ \chi(M(e)), \ldots,  \chi(M(S_4)) ]$ can be deduced  by \Eq{O} and  \Eq{I}, where $ M(g)$ is a matrix representation of the $g \in O_h$. 
By studying the action of $g \in O_h$ on 
  fermions, and hermitian and anti-hermitian parts of the bosonic link matrices,  the  character
multiplets  can be obtained as  
\beq
&&\chi^{\rm fermions} =   [8,2,0,0,0,   0, 0, 0, 4, 0] = A_{1g} \oplus  T_{2g} \oplus T_{1u}  \oplus 
A_{2u} \cr \cr
&&\chi^{\Re(\rm bosons)} =   [3,0,3,1,1, 3,0,  3, 1, 1]  = A_{1g} \oplus  E_g \cr \cr
&&\chi^{\Im(\rm bosons)} =   [3,0,-1,-1,1, -3,0,  1, -1, 1] = T_{2u}
\eqn{Ohsplitting}
\eeq
The gauge boson  remains irreducible under $O_h \subset SO(3)'$ and fills in 
the three dimensional  $T_{2u}$ representation. The scalars are as well in the three dimensional vector representation of twisted $SO(3)'$ group, however, they  are pseudo-vector  as opposed to being vectors. 
The characters for the inversion operation  are 
 $\chi^{\Im(\rm bosons)}(i)=-3$ and  $\chi^{\Re(\rm bosons)}(i)=+3$ 
 reflecting vector and pseudo-vector nature of these fields. 
   The pseudo-vector representation of  $SO(3)'$  is reducible under the $O_h$ subgroup, and splits as  $A_{1g} \oplus  E_g$.   
     For fermions,  everything works out beautifully. 
The irreducible representations  of the 
$O_h$  map into the  Dirac-K\"ahler twisted version of continuum: 
\beq
A_{1g} \oplus  T_{2g} \oplus T_{1u}  \oplus 
A_{2u}   \sim  \psi^{(0)} \oplus  \psi^{(1)} \oplus  \psi^{(2)} \oplus   \psi^{(3)} \sim 
   1 \oplus 3 \oplus 3  \oplus 1
\eeq

Let us reiterate the conclusion of the previous section: 
Although 
there is a one to one map between the irreducible representation of $O_h$ and Dirac-K\"ahler decomposition,  the gauging of the link fermions and $p$-form fermions are different. Thus, there is no gauge co-variant   identification of the 
various $p$-form lattice  fermions  and link formulation fermions  at  any finite lattice spacing. 
Of course, in the  continuum, the discrepancy disappears. This is the sense in which 
 the link fermion approach is tied with the  Dirac-K\"ahler structure of the continuum formulation.

\subsection{A $\CQ=0$ deformed matrix  model for link(2) formulations}
\label{nsmatrix}
The equality of the number of supersymmetries in the deformed matrix models and supersymmetric orbifold lattices  suggests that there must also exist a $\CQ=0$ deformed 
matrix models  which reproduce the non-supersymmetric     $\CQ=0$ link(2) lattices. The $\CQ=0$ matrix model for $\CN=(2,2)$ target theory  may be found by adapting  the techniques of the Ref.\cite{Unsal:2005us}.   The 
corresponding non-supersymmetric  deformed matrix model action is\beq
&&S^{\rm deformed}=\frac{1}{g^2} \Tr  \Bigg[ \half \Big(  [\bar z_1,z_1] + [\bar z_2,z_2]\Big)^2  
+2 \Big| e^{i \beta/2} z_1 z_2 - e^{-i \beta/2} z_2 z_1  \Big|^2   
 \cr  
&&  +  \sqrt 2 \left(   \alpha_{12}  [\bar z_1,  \alpha_{1\bar 2} ]_{\beta/4} +  \alpha_{12} [\bar z_2,  \alpha_{\bar 1 2} ]_{-\beta/4}   +  \alpha_{\bar 1 \bar 2}  [ z_1,  \alpha_{\bar 1  2} ]_{\beta/4}  - \alpha_{\bar 1\bar 2} [ z_2,  \alpha_{1 \bar 2} ]_{-\beta/4} 
\right)  \Bigg] \qquad 
\eqn{nsdef}
\eeq 
where  
\beq
 [\bar z_1,  \alpha_{1\bar 2} ]_{\beta/4} \equiv  e^{i{\beta/4}}  \;  \bar z_1  \alpha_{1\bar 2}  - 
 e^{- i{\beta/4}}  \;   \alpha_{1\bar 2}   \bar z_1 
\eeq

For  $\beta=0$,  the action is the dimensional reduction of the $d=4$ $\CN=1$ SYM theory down to $d=0$ and possesses   $\CQ=4$ supersymmetries. 
For $\beta \neq 0$, the  \Eq{nsdef} possesses no fermionic symmetry at all. This can be seen 
  explicitly by computation. For example, 
\beq
&& Q  z_1 = \sqrt 2  \alpha_{1 \bar2} \qquad \qquad  \qquad  \qquad \qquad Q \bar z_1 =  0 \cr 
&& Q  z_2 = \sqrt 2  \alpha_{\bar 1 2}  \qquad \qquad \qquad \qquad   \qquad Q \bar z_2=0 \cr
&& Q \alpha_{1 2} = - [\bar z_1, z_1]-  
 [\bar z_2, z_2] \qquad \qquad  \;\;  Q  \alpha_{1 \bar2} =0 \cr
 && Q \alpha_{\bar 1 \bar 2} = 2 [\bar z_1, \bar z_2] \qquad  \qquad  \qquad \qquad \;\;\;\;
  Q  \alpha_{\bar 1 2} = 0 .   
\eeq   
is an  on-shell supersymmetry of the  undeformed theory, the one given in    
Ref.\cite{Cohen:2003xe}.  But this is not a supersymmetry of the deformed  action given in  \Eq{nsdef}.   This is also true for all four supersymmetries or any linear combination thereof. 

It is in fact transparent  that the \Eq{nsdef} cannot have any of the fermionic symmetries of the  undeformed  theory. One reason is the mismatch of the $\beta$ commutators in the 
bosonic and fermionic parts of the action. 
The form of the deformed action in the fermionic   terms is 
\beq
S_f \sim    \alpha_{12} [ \bar z_1,   \alpha_{1\bar 2} ]_{\beta/4} = 
 e^{i{\beta/4}}   \alpha_{12} \bar z_1  \alpha_{1\bar 2}  - 
 e^{- i{\beta/4}}   \alpha_{12}  \alpha_{1\bar 2}   \bar z_1, \qquad 
\eeq
whereas, for example, the second bosonic term is 
\beq
S_b \sim  |[z_1, z_2]_{\beta/2}|^2 =   z_1z_2 \bar z_2 \bar z_1 +  z_2 z_1 \bar z_1 \bar z_2 -  e^{i{\beta}} z_1z_2 \bar z_1 \bar z_2 -  e^{-i{\beta}} z_2 z_1 \bar z_2 \bar z_1 
   \eeq
The variation of action under a supersymmetry transformation of the undeformed theory 
fails to vanish, because various terms which are supposed to cancel multiply different phase factors, for example $e^{\pm i \beta/4}$ versus $e^{\pm i \beta}$.   Thus, the action shown in 
\Eq{nsdef} 
is a $\CQ=0$ non-supersymmetric deformation of the $\CQ=4$ matrix model.  

\begin{figure}
\begin{center}
\includegraphics[angle=-90,width=5 in]{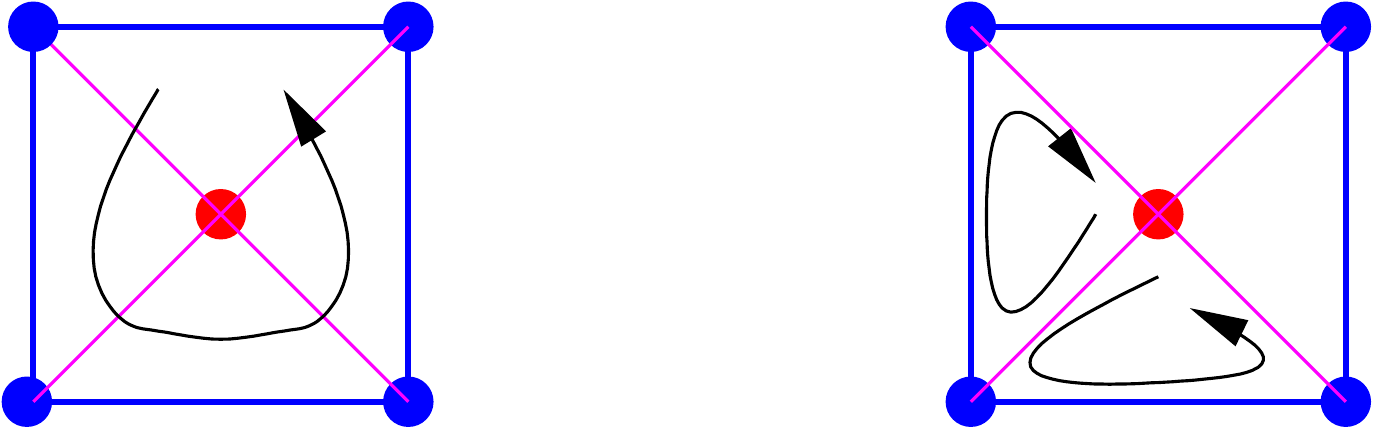}
\caption
    {\small
Examples of oriented square and triangular plaquettes.  If there is a background field, 
the net flux passing through the square is four times (in magnitude) the one of triangle. 
   If the theory is reduced to a single point
by using the twisted boundary conditions of 't Hooft, it produces the action with a 
$e^{\pm i \beta}$ phase for the reduction of square plaquettes, and $e^{\pm i \beta/4}$ phase for the triangle  plaquettes.    
 }
\label{linklat2}
\end{center}
\end{figure}

 There is also a nice physical interpretation for the difference of the various  phase factors as 
 discussed in detail in supersymmetric case in Ref.\cite{Unsal:2005us}. 
 The deformed matrix model \Eq{nsdef} can be obtained from the \Eq{link2a}, by dimensionally reducing the lattice action to a single point on the lattice by  using the 't Hooft's 
 twisted boundary conditions for lattice  fields. This is a  dimensional reduction on lattice 
 with a background flux.  In Fig.\ref{linklat} and  \Eq{link2a},  there are three types of plaquettes 
 that enters into the action, with  counter-clockwise and clock-wise orientations. These are 
  square plaquettes with area $4a^2$, triangular plaquettes with area $a^2$ and flipped-L plaquettes   with zero-area \cite{Unsal:2005yh}.    In the reduction with background flux, the phase factors appearing in the reduced matrix model are the fluxes passing through the corresponding surface prior to the reduction. See Fig.\ref{linklat2}. 
  These corresponds to the phases $e^{\pm i \beta/4}$,    $e^{\pm i \beta}$,
  for  triangular  and square plaquettes, and  identity otherwise.  \footnote{In orbifold lattices 
  (either supersymmetric or non-supersymmetric), the straight-forward  dimensional reduction to a single point  enhances the amount of supersymmetry to the level of parent matrix theory.  Recall that in continuum, there is no enhancement of number of supersymmetries upon dimensional reduction.   In lattice, the reduction by using the 't Hooft twisted boundary conditions
  however, keeps the number of the preserved supersymmetries intact. Both the lattice theory 
  and matrix model  has equal number of supersymmetries, which may be $\CQ={\rm few}$  
  \cite{Unsal:2005us}   or 
  $\CQ=0$ as in \Eq{nsdef}.}

 More specifically, consider the \Eq{nsdef} with $U(2Nk)$ algebra valued Grassmann odd and even matrices
 and $\beta=4 \frac{2 \pi}{2N}$. Both choices are for the convenience of the presentation. 
 Then, the background and fluctuations of the $\CQ=0$  matrix model can be transmuted into a lattice gauge theory on a $(2N)^2$ non-commutative lattice  with $U(k)$ gauge group.   The resulting action is  a familiar one, and gives
\beq 
S^{\rm deformed} = S^{\rm link(2)}|_{\star}. 
\eeq
where $S^{\rm link(2)}|_{\star}$ is same as in \Eq{link2a} with the modification of ordinary product into a  $\star$-product. 
As discussed in \S\ref{seiberg},  the commutative and non-commutative gauge theories 
carry equal amount of supersymmetries so long as no deformation in the Grassmann odd space 
is introduced as done in \cite{Seiberg:2003yz}.  Thus, $S^{\rm link(2)}|_{\star}$ has as much global  supersymmetries as $S^{\rm link(2)}$, which implies a $\CQ=0$ formulation.  
 \footnote{
Following footnote.\ref{QCD}, we may  substitute  group valued matrices instead of algebra valued complex matrices. The resulting theory is a matrix model regularization for $d=2$ dimensional QCD with adjoint fermions. The $d=4$ dimensional generalization is obvious. 
These are some exotic variations to the TEK models with commutative continuum limits.  The continuum limit can also be made non-commutative if desired.}

\section{Supersymmetric  lattice (SL)-twists and  topological field theories}

There are currently three types of proposal for a non-perturbative formulation of four dimensional 
$\CN=4$ SYM theory. These are 
\begin{itemize} 
\item{Exact Lattice supersymmetry [Orbifold,  geometric, Dirac-K\"ahler, topological  field theory motivated  Refs.\cite{ Kaplan:2005ta,Catterall:2005fd,Sugino:2004uv}] }
\item{Approaches with no  microscopic supersymmetry [with Ginsparg-Wilson 
fermions \cite{Elliott:2008jp},   
link(2)  of  Ref.\cite{D'Adda:2005zk} with the re-interpretation of  \S\ref{link2r}] }
\item{Supersymmetric/non-supersymmetric  deformed matrix models, Ref.\cite{Unsal:2005us} and generalization of  \S\ref{nsmatrix}}  
\end{itemize}

Although having a  non-perturbative definition of a supersymmetric gauge theory is important 
in its own right, it is expected that the broader applications of these lattices will be via gauge-gravity duality.
A lattice  definition of the various sixteen supercharge 
theories may open up a non-perturbative window into quantum gravity, string theory, and in the 
large $N$ limit into supergravity.   There are however two practical obstacles on the way:
the  fermion sign problem \footnote{This is a problem in the continuum, sourced by Yukawa couplings.  It is not possible to avoid it. } and the amount of  fine tuning. In four dimensional supersymmetric lattices, the amount of fine-tuning is still not fully understood, however, it is believed to be surmountable.  In $d \leq 3$, no fine tuning is necessary \cite{Kaplan:2002wv}.  If these theories can be solved  numerically, this will necessitate going beyond what is currently known in  supergravity, which is mostly limited to  two point functions  and thermal behavior as emphasized recently in \cite{Wiseman:2008qa}.  
In this section,  we will not make 
 remarks on the numerical investigations of supersymmetric theories which is already ongoing (See for example, \cite{Catterall:2007fp,Catterall:2008yz, Suzuki:2007jt}.).   
Rather, we wish   to address   where SL-twists fit within the class of all twisted 
supersymmetric theories, and their interrelation to topological theories.  

Recent studies  on lattice supersymmetry showed that 
supersymmetric lattices   in their  continuum limit,  always give the twisted
version of the supersymmetric theories. These are non-topological physical theories. However,
if desired, one can make them topological by declaring the scalar supercharge $Q$ as a BRST operator, and consider only the states $|\Omega\rangle$ annihilated by $Q$ as physical, i.e., 
$Q|\Omega\rangle=0$, modulo those which can be written as  $|\Omega'\rangle \sim 
Q|\Omega''\rangle $.
 In this sense, there is also an intimate connection between the topological 
field theories and supersymmetric lattices.

A common tread in both topological field theory and lattice supersymmetry is the existence of a nil-potent scalar supercharge $Q$. However, although all supersymmetric lattices may correspond  to the   twisted version topological  field theories (in the above sense), the reverse statement is not true. Given a supersymmetric twist with a scalar supercharge, we are not guaranteed to have a (non-problematic) supersymmetric  lattice formulation.  Below, we  examine 
this in connection with  the twists of the $\CN=4$ SYM theory in $d=4$.

\subsection{Three   twists of $\CN=4$ SYM in $d=4$:  Why SL-twist is special?} 
\label{sec:SL}
The  
$\CN=4$ SYM theory in $d=4$  has three inequivalent twists, i.e, three inequivalent 
embedding of an $SU(2) \times SU(2) $ into $SU(4)_{\cal R}$ symmetry, 
 each of which results in  one or two scalar supersymmetries \cite{Vafa:1994tf}. These classes 
  are most easily described  by providing  the decomposition of ${\bf 4}$ of $SU(4)$ in \Eq{super}. \footnote{This categorization is given on pg.8 of  Ref.\cite{Vafa:1994tf}. However, the first class   there  must  be as above. } 
\begin{eqnarray}
&& i) \;\; ({\bf 2,1}) \oplus ({\bf 1, 2}), \qquad  ({\rm SL-twist})    \cr 
&&  ii)\;\; ({\bf 1, 2}) \oplus ({\bf 1, 2})  \cr 
&& iii) \;\;  ({\bf 1, 2})  \oplus 
 ({\bf 1,1}) \oplus ({\bf 1,1}) 
 \eqn{embed}
 \end{eqnarray}
 Under these embedding, the supercharges (and fermions) transform as 
 \beq 
i) \; \;  {\rm fermions} \; && \rightarrow ({\bf 1,1}) \oplus 
({\bf 2,2})
\oplus [({\bf 3,1}) \oplus ({\bf 1,3})]
\oplus  ({\bf 2,2})  \oplus ({\bf 1,1})   \cr \cr
&&  \rightarrow 
{\bf  1 \oplus  4 \oplus 6 \oplus  4  \oplus 1 }    \qquad  ({\rm SL-twist})  \;  \cr \cr
ii) \; \;  {\rm fermions} \; && \rightarrow 2 \times  \Big[  {\bf (1,1)  \oplus 
(2,2) 
\oplus (3,1) } \Big] \cr \cr
iii) \; \;  {\rm fermions} \; && \rightarrow   \Big[ {\bf (1,1)  \oplus 
(2,2) 
\oplus (3,1)} \Big]  \oplus 2 \times \Big[ ({\bf 2,1}) \oplus ({\bf 1, 2}) \Big]
\eeq
under the   twisted rotation group 
\beq
 [SU(2)_L \times SU(2)_R ]'  \times (G_a)   \subset  [SU(2)_L \times SU(2)_R ] \times  
 SU(4)_{\cal R}
\eeq
where  class-dependent global ${\cal R}$-symmetry factor   $G_a$ $(a=i, ii, iii)$  is not important for our purpose. The gauge boson, which is a $SU(4)$  singlet,   transforms as 
$(\bf 2, 2)$.   The   scalars are singlet under the Lorentz symmetry and is in 
 ${\bf 6}={\bf  4 }\wedge {\bf 4} $, anti-symmetric representation  of $SU(4)$. 
 Therefore, \Eq{embed} 
 uniquely fixes  the decomposition  of     ${\bf 6}$ under the twisted rotation group, for example,  
\beq
i)\;\;  [   ({\bf 2,1}) \oplus ({\bf 1, 2})  ]  \wedge [   ({\bf 2,1}) \oplus ({\bf 1, 2})  ] 
= ({\bf 2}, {\bf 2}) \oplus 2({\bf 1,1}),   
\eeq
and similarly, 
\begin{eqnarray}
  ii)\;\; 3({\bf 1,1} ) \oplus ({\bf 1, 3})   \qquad 
 iii) \;\; 2 ( {\bf1,  2})  \oplus   2 (1,1)  
 \end{eqnarray}

As stated above, all three  twists support the existence of  at least one 
 nil-potent scalar supercharge $Q \sim  ({\bf 1,1}) $, with $Q^2 =0$, modulo gauge rotations. 
 The first two has two and the last has one.
  One would naively expect that, since 
 $Q^2=0$ does not interfere with any translation, it should be implementable on the lattice. 
 This intuition is not completely correct.   The reason is, what is allowed and what is not in a lattice regularization of supersymmetric theories has a number of other   constraints.  The existence of 
a nil-potent supercharge $Q$ in twisted version is not sufficient. 
 
First, note that, all three twists  have  a copy of the twist of $\CN=2$ SYM theory in 
$d=4$    \cite{Witten:1988ze} where 
eight   supercharges decompose as $   {\bf (1,1)  \oplus  (2,2) 
\oplus (3,1)} 
$. This structure exists in  a $L' \leftrightarrow R'$ symmetric manner in the 
first twist and asymmetric for the last two.  This means that, in case $i)$, instead of 
self-dual two-forms, we can just think of two-forms, without self-duality condition. 
 In lattice gauge theory, the implementation of 
the self-duality condition in a manifestly gauge covariant fashion is problematic.   
For example, in continuum, we will have $Q \psi^{\mu \nu, +}  =  F^{\mu \nu, +} \equiv  
 F^{\mu \nu} + 
\half \epsilon^{\mu \nu \rho \sigma}F_{\rho \sigma} $ where  both of 
$\psi^{\mu \nu, +}$ and $ F^{\mu \nu, +} $ are  in self-dual ${\bf (3,1)}$ representation.  
 The gauge-covariant implementation of the right-hand side on the lattice is not clear, and hence, the meaning of the left hand side (a self-dual Grassmann) is also unclear. This means, 
 the twist $ii)$ and   $iii)$ are not very pleasant from lattice point of view.  
Furthermore,  the  $iii)$ case also involves   
double-valued representation scalars and spinors, 
which are again in double-valued spinor representations of the lattice point group symmetry and 
do not have a natural habitat on  lattice, unlike the $p$-form $p$-cell mapping. 

The supersymmetric (orbifold) lattices always produce  the twists which do not involve 
any self-duality conditions.  All the   fields are in  single-valued integer spin representations, and naturally yields twist {\it i)}, with $Q(u,v) =Q(1,0) = Q$  or the dual 
$Q(u,v) = Q(0,1)= *Q^{(4)}$, in the notation of \Eq{uv}. 
   

{\bf Remark:} In the topological field theory literature,   it is sometimes 
asserted that the  twist of type $i)$ did not have any application  to physics up until the recent discussion of the dualities  of Ref.\cite{Kapustin:2006pk}. Most likely, what is meant here is topological applications. This twist had beautiful realizations and applications in  supersymmetric lattices. 
  Moreover, 
if we move to the application outside the supersymmetric or topological context, we immediately realize that  the twist of type $i)$ had the most application of all, in particular in  lattice gauge theory.
   The staggered fermions is the twisted version of the complex representation fermions,  and the reduced staggered fermions are the twisting applied to real representation fermions. Both are  used practically in numerical QCD and discussed in standard textbooks. However, the language is slightly different. 
In the next section, I will rephrase the staggered fermions  as twisting applied to  QCD.

\section{Twisting in QCD and staggered fermions} 
\label{sec:twisting}
  In this section, 
  I  rephrase the (reduced) staggered  fermions as  an elegant  application of twisting into QCD. 
   The main point of this  section  is shown  in Fig.\ref{godd}.  Needless to say, these theories 
   do not admit a topological interpretation, and the  twisted theory is necessarily physical.  (Recall that in supersymmetric context, we are free to make that choice and switch between the two.)   The discussion below borrows from  Refs.\cite{Rabin:1981qj,Becher:1982ud}  and the   lecture notes  \cite{Kaplan:2007zz}. In particular, the interrelation between 
    reduced staggered fermions and the 
    twists useful in the supersymmetric gauge theories was emphasized 
    to me by D.B. Kaplan. 

\subsection{Staggered fermions as  twisted complex representation fermions}
Consider massless  QCD on $\R^4$ with $N_f=4$ complex (for example, fundamental) representation fermions.    We label Dirac spinors as 
 \beq
\Psi_I= 
\left( \begin{array}{c}
     \psi_{\alpha }  \\
  \mybar \chi_{\dot \alpha}
       \end{array} \right)_I 
\eeq
and $\mybar \Psi_I$ where $I=1, \ldots, 4 $ is the flavor index.  The theory possesses  
\beq 
 G_{\rm QCD} = [SU(2)_L \times  SU(2)_R]_{\rm Lorentz} \times 
  [U(4)_{L} \times  U(4)_{R}]_{\rm flavor} 
\eeq 
space-time and (classical) chiral  flavor symmetries.
Under $G_{\rm QCD}$, the fermions transform as 
 \beq 
    \psi_{\alpha, I}   \sim ({\bf 2,1, 4,1}) , \; \; \;  \mybar \chi_{ \dot \alpha, I} \sim  
    ({\bf 1,2, 1, 4})
    \eeq
It is  convenient to use the vector-like  sum $U(4)_V \sim U(4)_{L+R}$  of  the flavor symmetry for our purpose. The Weyl components     $\psi_I$ ($\mybar \chi_I$)   fill in $({\bf 2,1, 4})$ and $({\bf 1,2, 4})$ under  $[SU(2)_L \times  SU(2)_R]_{\rm Lorentz} \times 
  [U(4)_{V}]_{\rm flavor}$.  
Clearly, the  flavor symmetries are sufficiently 
large  such that it can accommodate  a copy of  Lorentz group, i.e,  $SO(4) \subset U(4)_V$. 
Thus, the problem maps into the  discussion in \S\ref{td}, with twice as much fermion content.
The next steps are identical.  We use the decomposition of ${\bf 4}$ of $U(4)_{ V}$ into  
$({\bf 2,1})   \oplus  ({\bf 1,2})$ and perform a  diagonal embedding 
\beq 
SO(4)' \sim {\rm Diag} \; \Big( SO(4)_{\rm Lorentz} \times  SO(4)_{\rm flavor} 
\Big) \; .
\eeq
Under $SO(4)' $, the fermions of the original theory map into integer-spin representations, 
$p$-forms.  In an hyper-cubic lattice whose point group symmetry is a discrete subgroup of 
$G_{\rm point} \subset SO(4)'$,  it is natural to associate a $p$-form with a $p$-cell. 
This is the  Dirac-K\"ahler or geometric representation of fermions in lattice. 

The above procedure can equivalently be described as follows:  $\Psi_{\Upsilon I}$ is a four by four 
matrix, where $\Upsilon=1, \ldots 4$  is the Dirac spinor index and $I=1, \ldots 4$ is the flavor index. By using four dimensional Euclidean Dirac-matrices  $\gamma_{\mu}, 
 \mu=1, \ldots 4$,  
we can define a basis  ${\cal B} $ for $GL(4, \mathbb C)$ given by 
\beq 
{\cal B}\equiv \{ \Gamma^A , A=1, \ldots 16 \} = 
\{1, \gamma_{\mu},  \gamma_{[\mu\nu]},  \gamma_{[\mu\nu \rho]},   \gamma_{[\mu\nu \rho \sigma]} \} 
 \eeq
 where $[\ldots]$ denotes (normalized) anti-symmetrization.  
  ${\cal B}$ forms an orthonormal, complete basis for the $GL(4, \mathbb C)$.  The generators satisfy 
 $\Tr \Gamma^A \Gamma^B = 4  \delta^{AB}$. 
  Let us also define
 \beq 
 \{ \psi^A , A=1, \ldots 16 \} \equiv 
\{\psi, \psi_{\mu},  \psi_{[\mu\nu]},  \psi_{[\mu\nu \rho]},   \psi_{[\mu\nu \rho \sigma]} \} 
 \eeq
 the collection of $p$-form Grassmann valued fields. 
 Thus, we can write 
 \beq
  (\Psi)_{\Upsilon I} &&=  \sum_{A=1}^{16} \Gamma^A \psi^A , \qquad  \psi^A = \frac{1}{4} \Tr[  \Psi \Gamma^A]
  \eeq 
  or making the $SO(4)' $ transformation properties transparent 
  \beq 
 (\Psi)_{\Upsilon I}  && =   
   \left(\psi 1 + \psi^{\mu} \gamma_{\mu} + \frac{1}{2!} \psi^{\mu \nu}  \gamma_{[\mu\nu]} +
    \frac{1}{3!}  \psi^{\mu \nu \rho}  \gamma_{[\mu\nu \rho]} +  \frac{1}{4!} \psi^{\mu \nu \rho \sigma}  \gamma_{[\mu\nu \rho \sigma]} 
    \right)_{\Upsilon I}  \nonumber \\
       ( \mybar \Psi)_{\Upsilon I} && = \left(\mybar \psi 1 + \mybar \psi^{\mu} \gamma_{\mu} +  \frac{1}{2!}  \mybar  \psi^{\mu \nu}  \gamma_{[\mu\nu]} +  \frac{1}{3!} \mybar  \psi^{\mu \nu \rho}  \gamma_{[\mu\nu \rho]} +  \frac{1}{4!} \mybar \psi^{\mu \nu \rho \sigma}  \gamma_{[\mu\nu \rho \sigma]} 
    \right)_{\Upsilon I} 
    \eqn{Diracs}
 \eeq 
 Apparently,   the fermions transform 
 as $p$-form  integer spin representation of    $SO(4)' $, and on the lattice, they are naturally associated with the $p$-cells, sites, links, faces, cubes, and hypercubes.

 The equivalence between the geometric $p$-form fermions and 
 staggered fermions in ungauged lattice  theories is  well-know.  (The gauging slightly complicates things, we will not go into this detail.)  
 The $p$-form fermions can be mapped onto a lattice 
 with half the spacing as follows: Let  $\bfe_{\mu}, \mu=1, \ldots, 4$ denote  the four dimensional unit vectors, i.e., $(\bfe_{\mu})_{\nu} = \delta_{\mu \nu}$.
 The mapping takes site,  link, face, 3-cell and 4-cell 
  fermions into  $(0, 0, 0, 0)$,  $\bfe_{\mu}$,   $\bfe_{\mu}+   \bfe_{\nu}, \mu \neq \nu$, etc. That is, the fermions are mapped onto the sixteen corners of a hypercube.  The lattice periodicity is 
  $2 \bfe_{\mu}$ in each direction, twice the lattice spacing, as it is always the case with 
  staggered fermions. Each site carries two  Grassmann valued fields, one barred 
  and one unbarred.

\subsection{Reduced staggered fermions as  twisted real representation   fermions}
If there are $N_f$ massless Majorana fermions  in  a real representation of the gauge group such as  adjoint, then the flavor symmetry of the theory is $SU(N_f)_{\rm flavor}$.
  For example, for  QCD with four adjoint fermions [QCD(adj)], the (classical) symmetries of the theory are 
\beq 
 G_{\rm QCD(adj)} = [SU(2)_L \times  SU(2)_R]_{\rm Lorentz} \times 
  [SU(4)]_{\rm flavor} \times U(1)_A
\eeq 
Note that apart from the anomalous $U(1)_A$ factor, this is also the spacetime and flavor symmetry of the $\CN=4$ SYM, where flavor symmetry is called an ${\cal R}$-symmetry: 
\beq 
 G_{ N_f=4 \rm \; QCD(adj)} =  G_{\CN=4 \; {\rm SYM}}
\eeq
up to (unimportant) discrete symmetries. 
Consequently, (and not surprisingly), the reduced staggered 
fermions realization of the QCD(adj) is intimately related to the famous twist $i)$ of $\CN=4$ 
SYM \cite{Marcus:1995mq, Unsal:2006qp, Kapustin:2006pk}. This twist is also the one which arises naturally in the supersymmetric orbifold lattice \cite{Kaplan:2005ta}.  

The fermionic matter content of QCD(adj) would be conveniently described by either a Weyl spinor or equivalently, a Majorana spinor:
 \beq
\Psi^{\rm Maj.}_I= 
\left( \begin{array}{c}
     \psi_{\alpha }  \\
  \mybar \psi_{\dot \alpha}
       \end{array} \right)_I 
\eeq 
which fills in 
 \beq 
    \psi_{\alpha, I}   \sim ({\bf 2, 1,  4}) , \; \; \;\;\;  \mybar \psi_{ \dot \alpha, I} \sim  ({\bf 1,  2, 
     \bar 4})
    \eeq
representation.  
The twisting procedure is same as before, where $SU(4)_{\rm flavor}$ replaces the diagonal 
$SU(4)_V$ of the previous section.  The final expression is, 
 \beq
   (\Psi^{\rm M})_{\Upsilon I} = \left(\lambda 1 + \psi^{\mu} \gamma_{\mu} + \xi^{\mu \nu}  \gamma_{[\mu\nu]} + \xi^{\mu \nu \rho}  \gamma_{[\mu\nu \rho]} + \psi^{\mu \nu \rho \sigma}  \gamma_{[\mu\nu \rho \sigma]} 
    \right)_{\Upsilon I}
     \eqn{Weyls}
 \eeq 
where  the same notation as in the  $\CN=4$ SYM theory is used to ease the comparison. 
One can map the geometric $p$-form fermions (which come without unbarred fields) into the 
  {\it reduced  staggered fermions} as in the previous section. Note that  reduced staggered 
  fermions  has half as much degree of freedom on the lattice relative to the 
Kogut-Susskind staggered fermions.  This is simply, in the continuum, it corresponds to four flavors 
of Weyl (or Majorana) spinors  as opposed to the four flavors of Dirac  spinors. (Compare 
\Eq{Diracs} and \Eq{Weyls}.)

The  cubic supersymmetric orbifold lattices are natural  realization of 
$p$-form Dirac-K\"ahler fermions and  so is  the geometric formulation. See for example, \cite{Cohen:2003xe,Catterall:2004np}. During the writing of \cite{Unsal:2006qp}, it was 
not fully clear to me what the relation was between these two and the work of Sugino  
\cite{Sugino:2004uv},
 who
used   reduced staggered fermions in his constructions. The above symmetry arguments also 
clarify this point.  Indeed,  Ref.\cite{Takimi:2007nn, Damgaard:2007xi} recently 
constructed a  mapping  between these formulations at a finite lattice spacing.

\section{Discussion}
In even space-time dimensions, the deformed matrix models provide an alternative 
non-perturbative regularization for extended supersymmetric gauge theories, 
such as $\CN=4$ SYM in $d=4$ and  $\CN=(2,2)$ SYM in $d=2$ \cite{Unsal:2005us}.  These constructions 
are different from supersymmetric (orbifold) lattices\cite{Cohen:2003xe},  
and in particular, the $\beta$-flux 
deformation cannot be used to fully regularize  odd dimensional target theories in a Euclidean setting.  However,  the matrix model regularization also works for a curved space $S^2$, which  cannot be obtained via orbifold projections.

Our work also  provides a newer interpretation for the link approach \cite{D'Adda:2005zk, D'Adda:2007ax}  to lattice supersymmetry, by benefitting from Ref.\cite{Damgaard:2007eh}. 
 First, 
we classified the link approach lattices as link(1) [or orbifold(1)] and link(2) [or orbifold(2)].  
The first class preserves a (scalar) subset of supersymmetry and the latter 
preserves none. 
  Despite being non-supersymmetric theories at the cut-off, the link(2)/orbifold(2) class is quite intriguing. They have much larger point group symmetry $G_{\rm point}$ relative to the supersymmetry preserving formulations. In some sense, supersymmetry in the former is traded with the    large discrete chiral and discrete space-time 
symmetry   $G_{\rm point}$ realized in ${\rm Diag}(SO(d)_{\rm Lorentz} \times SO(d)_{\rm \cal R})$ in the latter.  In the classical continuum limit of the   link(2) type lattices,   one obtains 
 full extended  supersymmetry. 
 Whether this may be achieved in full quantum theory is not clear yet.  In particular, 
  there are  supersymmetry violating relevant operators, such as mass operators for scalars (which is forbidden in supersymmetric lattices), and it is important to check that higher dimensional operators  do not induce them radiatively.

We also demonstrated an equivalence. The supersymmetric deformed matrix models with 
$\CQ= {\rm few}$ produce a supersymmetric lattice gauge theory, both with a commutative and non-commutative continuum limit.  At finite lattice spacing, the non-commutative 
 lattice theories  have   identical actions  with the supersymmetric (orbifold) lattices, modulo 
the substitution of the  $\star$-product with the ordinary product of fields. Both formulations respect  the same set of supersymmetries.    
 The link(2) formulations can also be obtained from  deformed matrix models. The corresponding matrix models are new  $\CQ=0$  (non-supersymmetric) $\beta$ flux deformations of supersymmetric matrix models. This equivalence also confirms  that the is no exact supersymmetry associated with link(2) constructions.

There are also few other topics that we   either rephrased existing results in literature, or  we were unable to say much.
One is the generality  of the concept of twisting.  This is a useful notion in QCD via the use of 
 staggered fermions, in (non-topological) physical supersymmetric  theories,  in lattice gauge theory, and in topological supersymmetric field theories. 
  One direction   that can perhaps be improved 
  significantly is  the physical interpretation of the $\CN=\fourth$ SYM theory on $\R^4$, and its 
  lower dimensional counterparts.   
  

\acknowledgments
I thank Takemichi  Okui for his collaboration in the early stage of this work,
regarding in particular the structure of the Brillouin zones and \S\ref{SusySphere} in general.
I am thankful  to  David B. Kaplan  for many discussions on lattice fermions, and
 Simon Catterall,  Noboru Kawamoto, Kazuhiro Nagata  for valuable comments on the  draft. 
This work  is supported by the U.S.\ Department of Energy Grant DE-AC02-76SF00515.


\bibliography{emergentbib}

\providecommand{\href}[2]{#2}\begingroup\raggedright\begin{thebibliography}{10}

\bibitem{Witten:1988ze}
E.~Witten, {\it {Topological Quantum Field Theory}},  {\em Commun. Math. Phys.}
  {\bf 117} (1988) 353.

\bibitem{Kato:2005fj}
J.~Kato, N.~Kawamoto, and A.~Miyake, {\it {N = 4 twisted superspace from
  Dirac-Kaehler twist and off- shell SUSY invariant actions in four
  dimensions}},  {\em Nucl. Phys.} {\bf B721} (2005) 229--286,
  [\href{http://xxx.lanl.gov/abs/hep-th/0502119}{{\tt hep-th/0502119}}].

\bibitem{Leigh:1995ep}
R.~G. Leigh and M.~J. Strassler, {\it {Exactly marginal operators and duality
  in four-dimensional N=1 supersymmetric gauge theory}},  {\em Nucl. Phys.}
  {\bf B447} (1995) 95--136,
  [\href{http://xxx.lanl.gov/abs/hep-th/9503121}{{\tt hep-th/9503121}}].

\bibitem{Nishimura:2003tf}
J.~Nishimura, S.-J. Rey, and F.~Sugino, {\it {Supersymmetry on the
  noncommutative lattice}},  {\em JHEP} {\bf 02} (2003) 032,
  [\href{http://xxx.lanl.gov/abs/hep-lat/0301025}{{\tt hep-lat/0301025}}].

\bibitem{Unsal:2004cf}
M.~Unsal, {\it {Regularization of non-commutative SYM by orbifolds with
  discrete torsion and SL(2,Z) duality}},  {\em JHEP} {\bf 12} (2005) 033,
  [\href{http://xxx.lanl.gov/abs/hep-th/0409106}{{\tt hep-th/0409106}}].

\bibitem{Unsal:2005us}
M.~Unsal, {\it {Supersymmetric deformations of type IIB matrix model as matrix
  regularization of N = 4 SYM}},  {\em JHEP} {\bf 04} (2006) 002,
  [\href{http://xxx.lanl.gov/abs/hep-th/0510004}{{\tt hep-th/0510004}}].

\bibitem{Suzuki:2005dx}
H.~Suzuki and Y.~Taniguchi, {\it {Two-dimensional N = (2,2) super Yang-Mills
  theory on the lattice via dimensional reduction}},  {\em JHEP} {\bf 10}
  (2005) 082, [\href{http://xxx.lanl.gov/abs/hep-lat/0507019}{{\tt
  hep-lat/0507019}}].

\bibitem{Elliott:2008jp}
J.~W. Elliott, J.~Giedt, and G.~D. Moore, {\it {Lattice four-dimensional N=4
  SYM is practical}},  \href{http://xxx.lanl.gov/abs/0806.0013}{{\tt
  0806.0013}}.

\bibitem{lee-2007-76}
S.-S. Lee, {\it Emergence of supersymmetry at a critical point of a lattice
  model},  {\em Physical Review B} {\bf 76} (2007) 075103.

\bibitem{Kaplan:2002wv}
D.~B. Kaplan, E.~Katz, and M.~Unsal, {\it {Supersymmetry on a spatial
  lattice}},  {\em JHEP} {\bf 05} (2003) 037,
  [\href{http://xxx.lanl.gov/abs/hep-lat/0206019}{{\tt hep-lat/0206019}}].

\bibitem{Cohen:2003xe}
A.~G. Cohen, D.~B. Kaplan, E.~Katz, and M.~Unsal, {\it {Supersymmetry on a
  Euclidean spacetime lattice. I: A target theory with four supercharges}},
  {\em JHEP} {\bf 08} (2003) 024,
  [\href{http://xxx.lanl.gov/abs/hep-lat/0302017}{{\tt hep-lat/0302017}}].

\bibitem{Cohen:2003qw}
A.~G. Cohen, D.~B. Kaplan, E.~Katz, and M.~Unsal, {\it {Supersymmetry on a
  Euclidean spacetime lattice. II: Target theories with eight supercharges}},
  {\em JHEP} {\bf 12} (2003) 031,
  [\href{http://xxx.lanl.gov/abs/hep-lat/0307012}{{\tt hep-lat/0307012}}].

\bibitem{Endres:2006ic}
M.~G. Endres and D.~B. Kaplan, {\it {Lattice formulation of (2,2)
  supersymmetric gauge theories with matter fields}},  {\em JHEP} {\bf 10}
  (2006) 076, [\href{http://xxx.lanl.gov/abs/hep-lat/0604012}{{\tt
  hep-lat/0604012}}].

\bibitem{Onogi:2005cz}
T.~Onogi and T.~Takimi, {\it {Perturbative study of the supersymmetric lattice
  theory from matrix model}},  {\em Phys. Rev.} {\bf D72} (2005) 074504,
  [\href{http://xxx.lanl.gov/abs/hep-lat/0506014}{{\tt hep-lat/0506014}}].

\bibitem{Ohta:2006qz}
K.~Ohta and T.~Takimi, {\it {Lattice formulation of two dimensional topological
  field theory}},  {\em Prog. Theor. Phys.} {\bf 117} (2007) 317--345,
  [\href{http://xxx.lanl.gov/abs/hep-lat/0611011}{{\tt hep-lat/0611011}}].

\bibitem{Giedt:2003xr}
J.~Giedt, E.~Poppitz, and M.~Rozali, {\it {Deconstruction, lattice
  supersymmetry, anomalies and branes}},  {\em JHEP} {\bf 03} (2003) 035,
  [\href{http://xxx.lanl.gov/abs/hep-th/0301048}{{\tt hep-th/0301048}}].

\bibitem{Giedt:2006pd}
J.~Giedt, {\it {Deconstruction and other approaches to supersymmetric lattice
  field theories}},  {\em Int. J. Mod. Phys.} {\bf A21} (2006) 3039--3094,
  [\href{http://xxx.lanl.gov/abs/hep-lat/0602007}{{\tt hep-lat/0602007}}].

\bibitem{Unsal:2005yh}
M.~Unsal, {\it {Compact gauge fields for supersymmetric lattices}},  {\em JHEP}
  {\bf 11} (2005) 013, [\href{http://xxx.lanl.gov/abs/hep-lat/0504016}{{\tt
  hep-lat/0504016}}].

\bibitem{Catterall:2003wd}
S.~Catterall, {\it {Lattice supersymmetry and topological field theory}},  {\em
  JHEP} {\bf 05} (2003) 038,
  [\href{http://xxx.lanl.gov/abs/hep-lat/0301028}{{\tt hep-lat/0301028}}].

\bibitem{Catterall:2004np}
S.~Catterall, {\it {A geometrical approach to N = 2 super Yang-Mills theory on
  the two dimensional lattice}},  {\em JHEP} {\bf 11} (2004) 006,
  [\href{http://xxx.lanl.gov/abs/hep-lat/0410052}{{\tt hep-lat/0410052}}].

\bibitem{Catterall:2005fd}
S.~Catterall, {\it {Lattice formulation of N = 4 super Yang-Mills theory}},
  {\em JHEP} {\bf 06} (2005) 027,
  [\href{http://xxx.lanl.gov/abs/hep-lat/0503036}{{\tt hep-lat/0503036}}].

\bibitem{Catterall:2006jw}
S.~Catterall, {\it {Simulations of N = 2 super Yang-Mills theory in two
  dimensions}},  {\em JHEP} {\bf 03} (2006) 032,
  [\href{http://xxx.lanl.gov/abs/hep-lat/0602004}{{\tt hep-lat/0602004}}].

\bibitem{Sugino:2003yb}
F.~Sugino, {\it {A lattice formulation of super Yang-Mills theories with exact
  supersymmetry}},  {\em JHEP} {\bf 01} (2004) 015,
  [\href{http://xxx.lanl.gov/abs/hep-lat/0311021}{{\tt hep-lat/0311021}}].

\bibitem{Sugino:2004qd}
F.~Sugino, {\it {Super Yang-Mills theories on the two-dimensional lattice with
  exact supersymmetry}},  {\em JHEP} {\bf 03} (2004) 067,
  [\href{http://xxx.lanl.gov/abs/hep-lat/0401017}{{\tt hep-lat/0401017}}].

\bibitem{Sugino:2004uv}
F.~Sugino, {\it {Various super Yang-Mills theories with exact supersymmetry on
  the lattice}},  {\em JHEP} {\bf 01} (2005) 016,
  [\href{http://xxx.lanl.gov/abs/hep-lat/0410035}{{\tt hep-lat/0410035}}].

\bibitem{Sugino:2006uf}
F.~Sugino, {\it {Two-dimensional compact N = (2,2) lattice super Yang-Mills
  theory with exact supersymmetry}},  {\em Phys. Lett.} {\bf B635} (2006)
  218--224, [\href{http://xxx.lanl.gov/abs/hep-lat/0601024}{{\tt
  hep-lat/0601024}}].

\bibitem{Sugino:2008yp}
F.~Sugino, {\it {Lattice Formulation of Two-Dimensional N=(2,2) SQCD with Exact
  Supersymmetry}},  \href{http://xxx.lanl.gov/abs/0807.2683}{{\tt 0807.2683}}.

\bibitem{Unsal:2006qp}
M.~Unsal, {\it {Twisted supersymmetric gauge theories and orbifold lattices}},
  {\em JHEP} {\bf 10} (2006) 089,
  [\href{http://xxx.lanl.gov/abs/hep-th/0603046}{{\tt hep-th/0603046}}].

\bibitem{Catterall:2007kn}
S.~Catterall, {\it {From Twisted Supersymmetry to Orbifold Lattices}},  {\em
  JHEP} {\bf 01} (2008) 048, [\href{http://xxx.lanl.gov/abs/0712.2532}{{\tt
  0712.2532}}].

\bibitem{Takimi:2007nn}
T.~Takimi, {\it {Relationship between various supersymmetric lattice models}},
  {\em JHEP} {\bf 07} (2007) 010,
  [\href{http://xxx.lanl.gov/abs/0705.3831}{{\tt 0705.3831}}].

\bibitem{Damgaard:2007xi}
P.~H. Damgaard and S.~Matsuura, {\it {Relations among Supersymmetric Lattice
  Gauge Theories via Orbifolding}},  {\em JHEP} {\bf 08} (2007) 087,
  [\href{http://xxx.lanl.gov/abs/0706.3007}{{\tt 0706.3007}}].

\bibitem{Damgaard:2007eh}
P.~H. Damgaard and S.~Matsuura, {\it {Lattice Supersymmetry: Equivalence
  between the Link Approach and Orbifolding}},  {\em JHEP} {\bf 09} (2007) 097,
  [\href{http://xxx.lanl.gov/abs/0708.4129}{{\tt 0708.4129}}].

\bibitem{Damgaard:2008pa}
P.~H. Damgaard and S.~Matsuura, {\it {Geometry of Orbifolded Supersymmetric
  Lattice Gauge Theories}},  {\em Phys. Lett.} {\bf B661} (2008) 52--56,
  [\href{http://xxx.lanl.gov/abs/0801.2936}{{\tt 0801.2936}}].

\bibitem{Damgaard:2007be}
P.~H. Damgaard and S.~Matsuura, {\it {Classification of Supersymmetric Lattice
  Gauge Theories by Orbifolding}},  {\em JHEP} {\bf 07} (2007) 051,
  [\href{http://xxx.lanl.gov/abs/0704.2696}{{\tt 0704.2696}}].

\bibitem{Kaplan:2007zz}
D.~B. Kaplan, {\it {Supersymmetry on the lattice}},  {\em Eur. Phys. J. ST}
  {\bf 152} (2007) 89--112.

\bibitem{Elitzur:1982vh}
S.~Elitzur, E.~Rabinovici, and A.~Schwimmer, {\it {SUPERSYMMETRIC MODELS ON THE
  LATTICE}},  {\em Phys. Lett.} {\bf B119} (1982) 165.

\bibitem{D'Adda:2004jb}
A.~D'Adda, I.~Kanamori, N.~Kawamoto, and K.~Nagata, {\it {Twisted superspace on
  a lattice}},  {\em Nucl. Phys.} {\bf B707} (2005) 100--144,
  [\href{http://xxx.lanl.gov/abs/hep-lat/0406029}{{\tt hep-lat/0406029}}].

\bibitem{D'Adda:2004ia}
A.~D'Adda, I.~Kanamori, N.~Kawamoto, and K.~Nagata, {\it {Twisted N = 2 exact
  SUSY on the lattice for BF and Wess- Zumino}},  {\em Nucl. Phys. Proc.
  Suppl.} {\bf 140} (2005) 754--756,
  [\href{http://xxx.lanl.gov/abs/hep-lat/0409092}{{\tt hep-lat/0409092}}].

\bibitem{D'Adda:2005zk}
A.~D'Adda, I.~Kanamori, N.~Kawamoto, and K.~Nagata, {\it {Exact extended
  supersymmetry on a lattice: Twisted N = 2 super Yang-Mills in two
  dimensions}},  {\em Phys. Lett.} {\bf B633} (2006) 645--652,
  [\href{http://xxx.lanl.gov/abs/hep-lat/0507029}{{\tt hep-lat/0507029}}].

\bibitem{D'Adda:2007ax}
A.~D'Adda, I.~Kanamori, N.~Kawamoto, and K.~Nagata, {\it {Exact Extended
  Supersymmetry on a Lattice: Twisted N=4 Super Yang-Mills in Three
  Dimensions}},  {\em Nucl. Phys.} {\bf B798} (2008) 168--183,
  [\href{http://xxx.lanl.gov/abs/0707.3533}{{\tt 0707.3533}}].

\bibitem{Bruckmann:2006ub}
F.~Bruckmann and M.~de~Kok, {\it {Noncommutativity approach to supersymmetry on
  the lattice: SUSY quantum mechanics and an inconsistency}},  {\em Phys. Rev.}
  {\bf D73} (2006) 074511, [\href{http://xxx.lanl.gov/abs/hep-lat/0603003}{{\tt
  hep-lat/0603003}}].

\bibitem{Bruckmann:2006kb}
F.~Bruckmann, S.~Catterall, and M.~de~Kok, {\it {A critique of the link
  approach to exact lattice supersymmetry}},  {\em Phys. Rev.} {\bf D75} (2007)
  045016, [\href{http://xxx.lanl.gov/abs/hep-lat/0611001}{{\tt
  hep-lat/0611001}}].

\bibitem{Kaplan:2005ta}
D.~B. Kaplan and M.~Unsal, {\it {A Euclidean lattice construction of
  supersymmetric Yang- Mills theories with sixteen supercharges}},  {\em JHEP}
  {\bf 09} (2005) 042, [\href{http://xxx.lanl.gov/abs/hep-lat/0503039}{{\tt
  hep-lat/0503039}}].

\bibitem{Arianos:2007nv}
S.~Arianos, A.~D'Adda, N.~Kawamoto, and J.~Saito, {\it {Lattice supersymmetry
  in 1D with two supercharges}},  {\em PoS} {\bf LATTICE2007} (2007) 259,
  [\href{http://xxx.lanl.gov/abs/0710.0487}{{\tt 0710.0487}}].

\bibitem{Arianos:2008ai}
S.~Arianos, A.~D'Adda, A.~Feo, N.~Kawamoto, and J.~Saito, {\it {Matrix
  formulation of superspace on 1D lattice with two supercharges}},
  \href{http://xxx.lanl.gov/abs/0806.0686}{{\tt 0806.0686}}.

\bibitem{Vafa:1994tf}
C.~Vafa and E.~Witten, {\it {A Strong coupling test of S duality}},  {\em Nucl.
  Phys.} {\bf B431} (1994) 3--77,
  [\href{http://xxx.lanl.gov/abs/hep-th/9408074}{{\tt hep-th/9408074}}].

\bibitem{Andrews:2006aw}
R.~P. Andrews and N.~Dorey, {\it {Deconstruction of the Maldacena-Nunez
  compactification}},  {\em Nucl. Phys.} {\bf B751} (2006) 304--341,
  [\href{http://xxx.lanl.gov/abs/hep-th/0601098}{{\tt hep-th/0601098}}].

\bibitem{Ishii:2008ib}
T.~Ishii, G.~Ishiki, S.~Shimasaki, and A.~Tsuchiya, {\it {N=4 Super Yang-Mills
  from the Plane Wave Matrix Model}},
  \href{http://xxx.lanl.gov/abs/0807.2352}{{\tt 0807.2352}}.

\bibitem{Azeyanagi:2008bk}
T.~Azeyanagi, M.~Hanada, and T.~Hirata, {\it {On Matrix Model Formulations of
  Noncommutative Yang-Mills Theories}},
  \href{http://xxx.lanl.gov/abs/0806.3252}{{\tt 0806.3252}}.

\bibitem{Bietenholz:2006cz}
W.~Bietenholz, J.~Nishimura, Y.~Susaki, and J.~Volkholz, {\it {A
  non-perturbative study of 4d U(1) non-commutative gauge theory: The fate of
  one-loop instability}},  {\em JHEP} {\bf 10} (2006) 042,
  [\href{http://xxx.lanl.gov/abs/hep-th/0608072}{{\tt hep-th/0608072}}].

\bibitem{Marcus:1995mq}
N.~Marcus, {\it {The Other topological twisting of N=4 Yang-Mills}},  {\em
  Nucl. Phys.} {\bf B452} (1995) 331--345,
  [\href{http://xxx.lanl.gov/abs/hep-th/9506002}{{\tt hep-th/9506002}}].

\bibitem{Kapustin:2006pk}
A.~Kapustin and E.~Witten, {\it {Electric-magnetic duality and the geometric
  Langlands program}},  \href{http://xxx.lanl.gov/abs/hep-th/0604151}{{\tt
  hep-th/0604151}}.

\bibitem{Seiberg:2003yz}
N.~Seiberg, {\it {Noncommutative superspace, N = 1/2 supersymmetry, field
  theory and string theory}},  {\em JHEP} {\bf 06} (2003) 010,
  [\href{http://xxx.lanl.gov/abs/hep-th/0305248}{{\tt hep-th/0305248}}].

\bibitem{Nagata:2007mz}
K.~Nagata, {\it {On the Continuum and Lattice Formulations of N=4 D=3 Twisted
  Super Yang-Mills}},  {\em JHEP} {\bf 01} (2008) 041,
  [\href{http://xxx.lanl.gov/abs/0710.5689}{{\tt 0710.5689}}].

\bibitem{Nagata:2008zz}
K.~Nagata and Y.-S. Wu, {\it {Twisted SUSY Invariant Formulation of
  Chern-Simons Gauge Theory on a Lattice}},
  \href{http://xxx.lanl.gov/abs/0803.4339}{{\tt 0803.4339}}.

\bibitem{Nagata:2008xk}
K.~Nagata, {\it {Exact Lattice Supersymmetry at Large N}},
  \href{http://xxx.lanl.gov/abs/0805.4235}{{\tt 0805.4235}}.

\bibitem{Kovtun:2003hr}
P.~Kovtun, M.~Unsal, and L.~G. Yaffe, {\it {Non-perturbative equivalences among
  large N(c) gauge theories with adjoint and bifundamental matter fields}},
  {\em JHEP} {\bf 12} (2003) 034,
  [\href{http://xxx.lanl.gov/abs/hep-th/0311098}{{\tt hep-th/0311098}}].

\bibitem{Strassler:2003ht}
M.~J. Strassler, {\it {Non-supersymmetric theories with light scalar fields and
  large hierarchies}},  \href{http://xxx.lanl.gov/abs/hep-th/0309122}{{\tt
  hep-th/0309122}}.

\bibitem{Wiseman:2008qa}
T.~Wiseman and B.~Withers, {\it {Holographic renormalization for coincident
  Dp-branes}},  \href{http://xxx.lanl.gov/abs/0807.0755}{{\tt 0807.0755}}.

\bibitem{Catterall:2007fp}
S.~Catterall and T.~Wiseman, {\it {Towards lattice simulation of the gauge
  theory duals to black holes and hot strings}},  {\em JHEP} {\bf 12} (2007)
  104, [\href{http://xxx.lanl.gov/abs/0706.3518}{{\tt 0706.3518}}].

\bibitem{Catterall:2008yz}
S.~Catterall and T.~Wiseman, {\it {Black hole thermodynamics from simulations
  of lattice Yang-Mills theory}},
  \href{http://xxx.lanl.gov/abs/0803.4273}{{\tt 0803.4273}}.

\bibitem{Suzuki:2007jt}
H.~Suzuki, {\it {Two-dimensional $\mathcal{N}=(2,2)$ super Yang-Mills theory on
  computer}},  {\em JHEP} {\bf 09} (2007) 052,
  [\href{http://xxx.lanl.gov/abs/0706.1392}{{\tt 0706.1392}}].

\bibitem{Rabin:1981qj}
J.~M. Rabin, {\it {HOMOLOGY THEORY OF LATTICE FERMION DOUBLING}},  {\em Nucl.
  Phys.} {\bf B201} (1982) 315.

\bibitem{Becher:1982ud}
P.~Becher and H.~Joos, {\it {The Dirac-Kahler Equation and Fermions on the
  Lattice}},  {\em Zeit. Phys.} {\bf C15} (1982) 343.

\end{thebibliography}\endgroup
\bibliographystyle{JHEP} 
\end{document}